\newcommand{\mpfit}{MPFIT}
\newcommand{\ppxf}{pPXF}
\newcommand{\oiii}{[O\,\textsc{iii}]}
\newcommand{\nii}{[N\,\textsc{ii}]}
\newcommand{\sii}{[S\,\textsc{ii}]}
\newcommand{\oi}{[O\,\textsc{i}]}
\newcommand{\siii}{[S\,\textsc{iii}]}
\newcommand{\oii}{[O\,\textsc{ii}]}

\newcommand{\fevi}{[Fe\,\textsc{vi}]}
\newcommand{\fexiv}{[Fe\,\textsc{xiv}]}
\newcommand{\fevii}{[Fe\,\textsc{vii}]}
\newcommand{\fex}{[Fe\,\textsc{x}]}

\newcommand{\cloudy}{Cloudy}

\documentclass{aa}  
\usepackage{graphicx}
\usepackage{txfonts}
\usepackage[colorlinks=true,citecolor=blue,linkcolor=black]{hyperref}
\usepackage{color}

\begin{document} 

   \title{The MAGNUM survey: different gas properties in the outflowing and disc components in nearby active galaxies with MUSE}
  

   \author{M. Mingozzi\inst{1}\fnmsep\inst{2}
        \and
          G. Cresci\inst{2}
          \and
         G. Venturi\inst{3}\fnmsep\inst{2}
         \and
         A. Marconi\inst{3}\fnmsep\inst{2}
         \and
         F. Mannucci\inst{2}
         \and
        M. Perna\inst{2}
        \and
        F. Belfiore\inst{4}
        \and
        S. Carniani\inst{5}\fnmsep\inst{6}\fnmsep\inst{7}
        \and
      B. Balmaverde\inst{8}
       \and
      M. Brusa\inst{1}\fnmsep\inst{9}
       \and
      C. Cicone\inst{8}
       \and
     C. Feruglio\inst{10}
       \and
      A. Gallazzi\inst{2}
       \and
      V. Mainieri\inst{11}
       \and
      R. Maiolino\inst{5}\fnmsep\inst{6}
       \and
      T. Nagao\inst{12}
       \and
      E. Nardini\inst{2}
       \and
    E. Sani\inst{13}
       \and
      P. Tozzi\inst{2}
       \and
       S. Zibetti\inst{2}
          }

   \institute{Dipartimento di Fisica e Astronomia, Universit\`{a} degli Studi di Bologna, Via Piero Gobetti 93/2, I-40129, Bologna, Italy\\
              \email{matilde.mingozzi2@unibo.it}
         \and
             INAF -- Osservatorio Astrofisico di Arcetri, Largo E. Fermi 5, I-50157, Firenze, Italy
        \and
            Dipartimento di Fisica e Astronomia, Universit\`{a} degli Studi di Firenze, Via G. Sansone 1, I-50019 Sesto Fiorentino, Firenze, Italy
                 \and
         University of California Observatories - Lick Observatory, University of California Santa Cruz, 1156 High St., Santa Cruz, CA 95064, USA
        \and
            Cavendish Laboratory, University of Cambridge, 19 J. J. Thomson Ave., Cambridge, CB3 0HE, UK
        \and
            Kavli Institute for Cosmology, University of Cambridge, Madingley Road, Cambridge, CB3 0HA, UK
           \and
           Scuola Normale Superiore, Piazza dei Cavalieri 7, I-56126 Pisa, Italy  
      \and
         INAF -- Osservatorio Astronomico di Brera, Via Brera 28, I-20121, Milano, Italy
      \and
         INAF -- Osservatorio di Astrofisica e Scienza dello Spazio di Bologna, Via Piero Gobetti 93/3, I-40129, Bologna, Italy
          \and
             INAF -- Osservatorio Astronomico di Trieste, Via G.B. Tiepolo 11, I-34143 Trieste, Italy
           \and
             European Southern Observatory, Karl-Schwarzschild-Str. 2, 85748 Garching bei M\"unchen, Germany
          \and
             Research Center for Space and Cosmic Evolution, Ehime University, 2-5 Bunkyo-cho, Matsuyama 790-8577, Japan
          \and
             European Southern Observatory, Alonso de Cordova 3107, Casilla 19, Santiago 19001, Chile \\
             }

   \date{Received 03/10/2018; accepted 16/11/2018}

  \abstract
  {We investigated the interstellar medium (ISM) properties of the disc and outflowing gas in the central regions of nine nearby Seyfert galaxies, all characterised by prominent conical or biconical outflows. These objects are part of the Measuring Active Galactic Nuclei Under MUSE Microscope (MAGNUM) survey, which  aims to probe their physical conditions and ionisation mechanism by exploiting the unprecedented sensitivity of the Multi Unit Spectroscopic Explorer (MUSE), combined with its spatial and spectral coverage.
Specifically, we studied the different properties of the gas in the disc and in the outflow with spatially and kinematically resolved maps by dividing the strongest emission lines in velocity bins. We associated the core of the lines with the disc, consistent with the stellar velocity, and the redshifted and the blueshifted wings with the outflow. We measured the reddening, density, ionisation parameter, and dominant ionisation source of the emitting gas for both components in each galaxy. 
We find that the outflowing gas is characterised by higher values of density and ionisation parameter than the disc, which presents a higher dust extinction. Moreover, we distinguish high- and low-ionisation regions across the portion of spatially resolved narrow-line region (NLR) traced by the outflowing gas. The high-ionisation regions characterised by the lowest \nii/H$\alpha$ and \sii/H$\alpha$ line ratios generally trace the innermost parts along the axis of the emitting cones where the \siii/\sii \, line ratio is enhanced, while the low-ionisation regions follow the cone edges and/or the regions perpendicular to the axis of the outflows, also characterised  by a higher \oiii \,velocity dispersion. A possible scenario to explain these features relies on the presence of two distinct populations of line emitting clouds: one is optically thin to the radiation and is characterised by the highest excitation, while the other is  optically thick and is impinged by a filtered,  and thus harder, radiation field which generates strong low-excitation lines. The highest values of \nii/H$\alpha$ and \sii/H$\alpha$ line ratios may be due to shocks and/or a hard filtered radiation field from the active galactic nucleus.}


\titlerunning{Gas properties in the outflowing and disc components in nearby active galaxies with MUSE}
\authorrunning{M. Mingozzi et al.}
\maketitle
%

\section{Introduction}
The gas located in the inner kiloparsecs of active galaxies plays a crucial role in many important processes;  it fuels accretion onto the central black hole (BH) in active galactic nuclei (AGN) and absorbs the energy and momentum injected in the interstellar medium (ISM) by the AGN, becoming a key ingredient in the BH-galaxy co-evolution. The AGN radiation acts as a flashlight that illuminates and ionises the nearby gas, forming the narrow-line region (NLR). Outflows powered by accretion disc winds and relativistic jets interact with the NLR gas, possibly inducing feedback effects on the host galaxy \citep{cresci2018}. Feedback mechanisms appear necessary to explain the present-day scaling relations between the host galaxy properties and BH mass and prevent galaxies from overgrowing (e.g. \citealt{silk1998,fabian2012}).

The NLR can extend by several kpc in Seyfert galaxies (i.e. extended narrow-line regions, ENLRs, e.g. \citealt{greene2012,sun2017,sun2018}) and has a relatively low-density ($n_e\sim 10^2-10^4$~cm$^{-3}$, e.g. \citealt{vaona2012}). It comprises ionised and neutral gas set-up in clouds, emitting low- and high-ionisation lines,  such as \sii$\lambda$6717,31 and \oiii$\lambda\lambda$4959,5007, respectively,  of elements photoionised by the non-stellar continuum emission of the AGN (e.g. \citealt{netzer2015}). 
These clouds are often observed to be entrained in AGN-driven outflows (e.g. \citealt{hutchings1998, crenshaw_kraemer2000}) that, especially in nearby Type 2 AGN, are observed to be fan-shaped, showing clearly defined biconical structures (e.g. \citealt{storchi-bergmann1992, das2006, crenshaw2010, cresci2015b,venturi2018}). 
Therefore, resolved NLRs are ideal laboratories for  investigating in detail the effects of feeding and feedback, and for constraining the ionisation structure of the emitting regions.

Specifically, diagnostic diagrams constructed from emission-line intensity ratios are widely used to differentiate between excitation mechanisms. Firstly, \citet{baldwin1981} proposed the \nii/H$\alpha$ versus \oiii/H$\beta$ diagram in order to discriminate H~II-like sources from objects photoionised by a harder radiation field, such as a power-law continuum by an AGN or shock excitation. This classification scheme was enlarged and improved by \citet{veilleux1987}, who added the \sii/H$\alpha$ and the \oi/H$\alpha$ line ratios. These are known as  Baldwin-Phillips-Terlevich (BPT) diagrams and involve line ratios that are not significantly separated in wavelength, minimising the effects of differential reddening by dust, and can discriminate among three different groups of emission line galaxies: those excited by starbursts, the Seyfert galaxies, and those found in the low-ionisation (nuclear) emission-line regions (LI(N)ERs, \citealt{heckman1980, belfiore2016a}).

There have been many studies aimed at identifying the ionisation source in Seyfert and star-forming galaxies through the observed intensity ratios. Photoionisation models have had  remarkable success in reproducing high-ionisation AGN lines, such as \oiii\ and  \siii \,\citep{ferguson1997,komossa1997}, while the excitation mechanisms for low-ionisation lines (e.g. \nii, \sii, \oi) in the region overlapping with star-forming galaxies remains uncertain. Furthermore, fast radiative shocks produced by different phenomena (e.g. cloud-cloud collisions, the expansion of H II regions into the surrounding ISM, outflows from young stellar objects, supernova blast waves, outflows from active and starburst galaxies, and collisions between galaxies) can be a powerful ionising source and can provide an important component of the total energy budget, and in some circumstances dominate the line emission spectrum \citep{dopita1995, dopita1996, allen2008}. 

The recent development of powerful integral field unit (IFU) spectrographs has allowed us to make a considerable step forward in understanding galaxy evolution, providing spatially resolved information on the kinematics of the gas and stars, and on galaxy properties, in terms of gas ionisation and chemical abundances (e.g. the CALIFA, \citealt{sanchez2012}; SAMI, \citealt{allen2015}; and MaNGA, \citealt{bundy2015} surveys, locally, and \citealt{forster2009,law2009,cresci2010,contini2012,wisnioski2015,stott2016,turner2017} at high redshift). Integral field spectroscopy (IFS) is a powerful tool also for the study of outflows, both in the local Universe (e.g. \citealt{storchi-bergmann2010, sharp2010, riffel2013, mcelroy2015, cresci2015b, karouzos2016a, karouzos2016b}) and at higher redshift (e.g. \citealt{cano-diaz2012,liu2013,cresci2015,perna2015,harrison2016,carniani2016,forster2018}). 
Integral field data can indeed shed light on the impact of AGN activity on the host galaxy through the kinematic signatures visible in emission-line profiles and the spatial distribution of emission-line flux ratios. 

In this paper, we present our results obtained with the survey  Measuring Active Galactic Nuclei Under MUSE Microscope (MAGNUM, P.I. Marconi, \citealt{cresci2015b, venturi2017, venturi2018}, Venturi et al. 2018 in preparation) survey, which  studies nearby AGN ($D_{L} < 50$~Mpc) by probing the physical conditions of the NLRs/ENLRs, and also studies the interplay between nuclear activity and star formation (SF), and the properties of outflows thanks to the unprecedented combination of spatial and spectral coverage provided by the integral field spectrograph MUSE at VLT \citep{bacon2010}. 

Specifically, in this paper we use IFS to conduct detailed studies on how ISM properties (e.g. density, temperature, ionisation, reddening) and the relative contributions of different ionisation mechanisms vary within individual galaxies, all characterised by prominent conical or biconical outflows. Taking a step forward, we kinematically disentangle the disc and outflow components, and compare their gas properties.

The paper is structured as follows. In Sect. \ref{sec:sample} we introduce the galaxy sample and the data analysis, while in Sect. \ref{diskVSoutflow} we explain our approach to disentangle the outflow component from the disc, and we investigate the gas properties through proper diagnostics and the gas excitation through BPT diagrams. Finally, in Sect. \ref{sec:discussion} we discuss our results, making a comparison with photoionisation and shock models found in the literature, and in Sect. \ref{sec:conclusion} we conclude with a summary of the results.

\section{Galaxy sample}\label{sec:sample}
MAGNUM galaxies have been selected by cross-matching the optically selected AGN samples of \citet{maiolino1995} and \citet{risaliti1999}, and Swift-BAT 70-month Hard X-ray Survey (\citealt{baumgartner2013}), choosing only sources observable from Paranal Observatory ($-70^\circ<\delta<20^\circ$) and with a luminosity distance $D_{\rm{L}}<50$~Mpc. 
In Venturi et al. (in preparation), we present our sample, explaining in detail the selection criteria, data reduction, and analysis, and investigating the kinematics of the ionised gas. In this work we present our results for nine Seyfert galaxies, namely Centaurus~A, Circinus, NGC~4945, NGC~1068, NGC~1365, NGC~1386, NGC~2992, NGC 4945, and NGC~5643. In the following, we briefly describe their main features. \\

\subsection{MAGNUM galaxies}\label{sec:sample1}
{\bf Centaurus~A} (Seyfert 2 - D$_{\rm{L}}$~$\sim 3.82$~Mpc - 1"~$\sim 18.5$~pc) is an early-type galaxy characterised by a major axis stellar component, a minor axis dust-lane, and a gas disc. The last follows the dust-lane and comprises the majority of the observed ionised, atomic, neutral, and molecular gas \citep{morganti2010}. It is characterised by the presence of a spectacular double-sided radio jet revealed in both  the radio and X-rays (e.g. \citealt{blanco1975,hardcastle2003}), which was found to match a diffuse highly ionised halo of \oiii \, \citep{kraft2008} first identified by \citet{bland2003}. 

{\bf Circinus} (Seyfert 2 - $D_{\rm{L}}\sim 4.2$~Mpc - 1"~$\sim 20.4$~pc) is a nearby gas rich spiral, hosting one of the nearest Seyfert 2 nuclei known.
The Seyfert 2 nature of Circinus is supported by the optical images showing a spectacular one-sided and wide-angled kpc-scale \oiii \, cone, first revealed by \citet{marconi1994} (see also \citealt{veilleux1997}), and the optical/IR spectra rich in prominent and narrow coronal lines \citep{oliva1994,moorwood1996}. Furthermore, from the X-ray spectral analysis Circinus is a highly obscured AGN, but with strong thermal dust emission dominated by heating from nuclear star formation \citep{matt2000}. 

{\bf IC~5063} (Seyfert 2 - $D_{\rm{L}}\sim 45.3$~Mpc - 1"~$\sim 219.6$~pc) is a massive early-type galaxy with a central gas disc and a radio jet that propagates for several hundred parsec before it extends beyond the disc plane \citep{morganti1998}. An extended ($\sim20$~kpc) \oiii-dominated double ionisation cone with an X morphology is observed \citep{colina1991}, ubiquitously dominated by photoionisation from the AGN \citep{sharp2010}.
	In this galaxy there is one of the most spectacular examples of jet-driven outflow, with similar features in the ionised, neutral atomic, and molecular phase, since the radio plasma jet is expanding into a clumpy gaseous medium, possibly creating a cocoon of shocked gas which is pushed away from the jet axis \citep{morganti1998,sharp2010,tadhunter2014,dasyra2015,oosterloo2017}. 

{\bf NGC~1068} (Seyfert 2 - $D_{\rm{L}}\sim$~12.5~Mpc - 1"~$\sim$60~pc) is one of the closest and archetypal Seyfert 2 galaxies. This galaxy also hosts  powerful starburst activity (e.g. \citealt{lester1987,bruhweiler1991}) and is characterised by a large-scale oval and a nuclear stellar bar aligned NE-SW \citep{scoville1988,thronson1989}. NGC 1068 is known to host a radio jet, observed from X-ray to radio wavelengths \citep{bland-hawthorn1997}. Its activity seems confined to bipolar cones centred  on the AGN, intersecting the plane of the disc with an inclination of $\sim 45^\circ $, such that the disc interstellar medium to the NE and SW sees the nuclear radiation field directly \citep{cecil1990,gallimore1994}. 
This galaxy shows  clear evidence of outflowing material in both the ionised (e.g. \citealt{macchetto1994,axon1997,cecil2000,barbosa2014}) and molecular gas \citep{burillo2014,gallimore2016}, possibly driven by the radio jet (\citealt{cecil1990,gallimore1994,axon1997}), located in the  north-east and south-west directions. 

{\bf NGC~1365} (Seyfert 1.8 - $D_{\rm{L}}\sim 18.6$~Mpc - 1"~$\sim 90.2$~pc) is a local spiral, characterised by a strong bar and prominent spiral structure, displaying nuclear activity and ongoing SF. A comprehensive review about the early works on this object is given by \citet{lindblad1999}.
Our MUSE data was already published by \citet{venturi2018}, who studied the circumnuclear gas in its different phases by comparing MUSE and Chandra data and analysing the nuclear and extended gas outflows. Specifically, \citet{venturi2018} revealed that \oiii \, emission that traces the kpc-scale biconical  outflow, ionised by the AGN, is nicely matched with the soft X-rays. However, the hard X-ray emission from the star-forming circumnuclear ring suggests that SF might in principle contribute to the outflow.

{\bf NGC~1386} (Seyfert 2 - $D_{\rm{L}}\sim 15.6$~Mpc - 1"~$\sim 76$~pc), classified as a Sb/c spiral galaxy, is one of nearest Seyfert 2 galaxies and has an inclination of $\sim65^\circ$ with respect to the line of sight. 
\citet{lena2015} analysed the gas and stellar kinematics in the inner regions ($530\times680$~pc), obtained from IFS GEMINI South telescope observations, revealing the presence of a bright nuclear component, two elongated structures to the north and south of the nucleus, and low-level emission extending over the whole field of view (FOV), probably photoionised by the AGN. Furthermore, \citet{lena2015} identified a bipolar outflow aligned with the radiation cone axis. This feature is resolved in HST observations \citep{ferruit2000} into two bright knots that represent the approaching and receding sides of the outflow, respectively. Finally, \citet{lena2015} suggested the presence of an outflow and/or rotation in a plane roughly perpendicular to the AGN radiation cone axis, extending 2"$-$3" on both sides of the nucleus, characterised by an enhanced velocity dispersion and coincident with a faint, bar-like emission structure revealed by HST imaging \citep{ferruit2000}.

{\bf NGC~2992} (Seyfert 1.9 - $D_{\rm{L}}\sim$~31.5~Mpc - 1"~$\sim$150~pc) is a spiral galaxy hosting a Seyfert nucleus and interacting with the companion galaxy NGC 2993, located to the south-east. This galaxy is classified as a Seyfert 1.9 in the optical, on the basis of a broad H$\alpha$ component with no corresponding H$\beta$ component in its nuclear spectrum, although its type has changed between 1.5 and 2 according to past observations \citep{trippe2008}. 
Looking at the optical image of this galaxy, a prominent dust lane extending along the major axis of the galaxy and crossing the nucleus stands out \citep{ward1980}.
Opposite ENLR cones extend along either side of the disc with large opening angles, and are both easily visible due to the high inclination of NGC 2992 (e.g. \citealt{allen1999}).
Evidence of a wide biconical  large-scale outflow, extending above and below the plane of the galaxy and possibly driven by AGN activity \citep{friedrich2010}, comes from H$\alpha$ and \oiii \,emission and soft X-ray observations (e.g. \citealt{colina1987,colbert1996a,colbert1996b,veilleux2001}).

{\bf NGC~4945} (Seyfert 2 - $D_{\rm{L}}\sim3.7$~Mpc - 1"~$\sim$18~pc) is an almost perfectly edge-on spiral galaxy, well known for being one of the closest objects where AGN and starburst activity coexist. Its central region is characterised by a very strong obscuration, associated with a prominent dust lane aligned along the major axis of the galactic disc, revealed by near- and mid-infrared spectroscopy typical of highly obscured ultra-luminous infrared galaxies (ULIRGs) \citep{beaupuits2011}.
This galaxy is characterised by a biconical outflow, clearly visible from the \nii \, emission map shown in \citet{venturi2017}, where the NW lobe is far brighter than the SE lobe, which is observed only at 15" from the centre, likely being completely dust-obscured at smaller radii. The AGN presence is only supported by X-rays observations (e.g. \citealt{guainazzi2000}), meaning that either its UV radiation is totally obscured along all lines of sight or that it lacks UV photons with respect to X-rays, implying in both cases a non-standard activity \citep{marconi2000}.

{\bf NGC~5643} (Seyfert 2 - $D_{\rm{L}}\sim17.3$~Mpc - 1"~$\sim$84~pc) is a barred almost face-on Seyfert 2. This galaxy is characterised by a well-known ionisation cone extending eastward of the nucleus and parallel to the bar (e.g. \citealt{schmitt1994,simpson1997,fischer2013}), and by a sharp, straight dust lane extended from the sides of the central nucleus out to the end of the bar, roughly parallel to its major axis and clearly visible to the east of the nucleus \citep{ho2011}. Our MUSE data of this target has been already published by \citet{cresci2015b}, who analysed the double-sided ionisation cone, revealed as a blueshifted, asymmetric wing of the \oiii \, emission line, up to a projected velocity of $\sim - 450$~km~s$^{-1}$, parallel to the low-luminosity radio and X-ray jet, and possibly collimated by a dusty structure surrounding the nucleus. Furthermore, \citet{cresci2015b} found signs of positive feedback triggered by the outflowing gas in two star-forming clumps, located at the edge of the dust lane of the bar (see \citealt{silk2013}), and also characterised by a faint CO(2--1) emission \citep{alonso-herrero2018}. 

\subsection{Data analysis}\label{sec:analysis}
The data reduction was performed using the MUSE pipeline (v1.6). The final datacubes consist of $\sim 300 \times 300$ spaxels, for a total of over 90000 spectra with a spatial sampling of 0.2"~$\times$~0.2" and a spectral resolution 
going from 1750 at 4650 \AA \,to 3750 at 9300 \AA. The FOV of 1'~$\times$~1' covers the central part of the galaxies, spanning from 1 to 10 kpc, according to their distance. The average seeing of the observations, derived directly 
from foreground stars in the final datacubes, is $\sim0.8$". Here we summarise the main steps of the data analysis, while full details are given in the survey presentation paper by Venturi et al. (in preparation). 

The datacubes were analysed making use of a set of custom python scripts in order to subtract the stellar continuum and fit the emission lines with multiple Gaussian components where needed. First of all, the stellar continuum was 
modelled using a linear combination of \citet{vazdekis2010} synthetic spectral energy distributions for single stellar population models in the wavelength range $4750-7500$~\AA. In order to subtract the continuum around the \siii$\lambda$9069, we performed a polynomial fit to the local continuum since this line is not contaminated by underlying absorptions. We applied a Voronoi tessellation 
\citep{cappellari2003} in order to achieve an average signal-to-noise ratio (S/N) per wavelength channel of 50 on the continuum under 5530 \AA, we performed the continuum fit using the Penalized Pixel-Fitting (\ppxf; \citealt{cappellari2004}) code on the binned spaxels, to reproduce the systemic velocity and the velocity dispersion of the stellar absorption lines. To do this, we simultaneously fit the main emission lines included in the selected wavelength 
range (H$\beta$ - \oiii$\lambda\lambda$4959,5007 - \oi$\lambda\lambda$6300,64 - H$\alpha$ - \nii$\lambda\lambda$6548,84 - \sii$\lambda\lambda$6717,31) in order to better constrain the absorption underlying the Balmer lines. 
Fainter lines and regions affected by sky residuals were masked. Then, the fitted stellar continuum in each Voronoi bin was subtracted spaxel by spaxel. Since NGC~1365 and NGC~2992 host a Seyfert 1 nucleus, we made use of an 
additional template when fitting the very central bins  in order to  simultaneously reproduce the broad-line region (BLR) features (broad Balmer lines and Fe II forest) and the BH accretion disc continuum. 

From the continuum-subtracted cube, we generated both a spatially smoothed data cube and a Voronoi-binned data cube. The former is obtained by spatially smoothing the cube with a Gaussian kernel having $\sigma = 1$~spaxel (i.e. 0.2''), without 
degrading significantly the spatial resolution, given the observational seeing. 
The latter, produced by applying a Voronoi binning in the range $\sim4850-4870$~\AA \, in the vicinity of H$\beta$ and requiring an average S/N per wavelength channel of at least 3, is used to make comparisons among the galaxies 
(see Sect.~\ref{sec:discussion}).
Finally, in each spaxel, we fit the gas emission lines in the same spectral range used for the stellar fitting, making use of \mpfit \, \citep{markwardt2009}, with one, two, three, or four Gaussians according to the peculiarities of the line profile, tying the 
velocity and widths of each component to be the same for all the lines and leaving the intensities free to vary, apart from the \oiii, \oi,  \,and \nii\, doublets, where an intrinsic ratio of $0.333$ between the two lines was used. We applied 
a reduced $\chi^2$ selection to the different fits made in each spaxel in order to define where multiple components were really needed to reproduce the observed spectral profiles, with the idea of keeping the number of fit parameters 
as low as possible and using multiple components only in case of complex non-Gaussian line profiles (for further details see Venturi et al. 2018 in preparation). This happens mainly in the central parts of the galaxies and in the outflowing cones. 

\section{Gas properties: disc versus outflow}\label{diskVSoutflow}
Excitation and ionisation conditions, and dust attenuation of the ISM in each MAGNUM galaxy can be investigated through specific emission-line ratios. Thanks to IFS, we explore the different properties of the disc and the outflow of our galaxies by separating and analysing the different gas components across the MUSE FOV.

Specifically, we define velocity channels of $\sim50$~km/s, ranging from $-1000$~km/s to $+1000$~km/s around the main emission lines (H$\beta$ - \oiii \, - \oi \, - H$\alpha$ - \nii \, - \sii \, - \siii), centring the zero velocity to the measured stellar velocity in each spaxel of the MUSE FOV for all the galaxies since the stellar velocity is generally a good approximation of the gas velocity in the disc.
In order to derive the flux in each velocity bin, we integrate the fitted line profile, performed as described in Sect.~\ref{sec:analysis}, within each velocity channel of the emission lines taken into account.
However, Centaurus~A is characterised by a strong misalignment between stars and gas (e.g. \citealt{morganti2010}), and thus we consider the global systemic velocity ($ v_{\rm{sys}}=547$~km/s) as a reference for the gas disc velocity. 

We associate the velocity channels close to the core of the lines in the fitted line profile with the disc (hereafter disc component), and the sum of the redshifted and blueshifted channels with the outflow (hereafter outflow component). Specifically, the disc component comprises velocity in the range $-100 \rm{~km/s} < v < 100 \rm{~km/s}$ for Circinus, NGC~1365, and NGC~1386, and $-150 \rm{~km/s} < v < 150 \rm{~km/s}$ for the other galaxies, while the outflow component is associated with  velocities $v > + 150$~km/s and $v < -150$~km/s for Circinus, NGC~1365 and NGC~1386, $v > + 250$~km/s and $v < -250$~km/s for NGC~1068, and $v > + 200$~km/s and $v < -200$~km/s for the other galaxies. These values are chosen after an accurate spaxel-by-spaxel analysis of the spectra in the outflowing regions of the galaxies. Overall, the disc component represents the low-velocity component of the ionised gas, which is rotating similarly to the stars, while the outflow component (i.e. the high-velocity component) is moving faster than the stellar velocity, and is partly blueshifted and partly redshifted with respect to it. Since we are looking at the central regions of AGN galaxies, the NLR is present in both the disc and in the outflow components.

\begin{figure}
   \centering
   \includegraphics[width=1.\columnwidth]{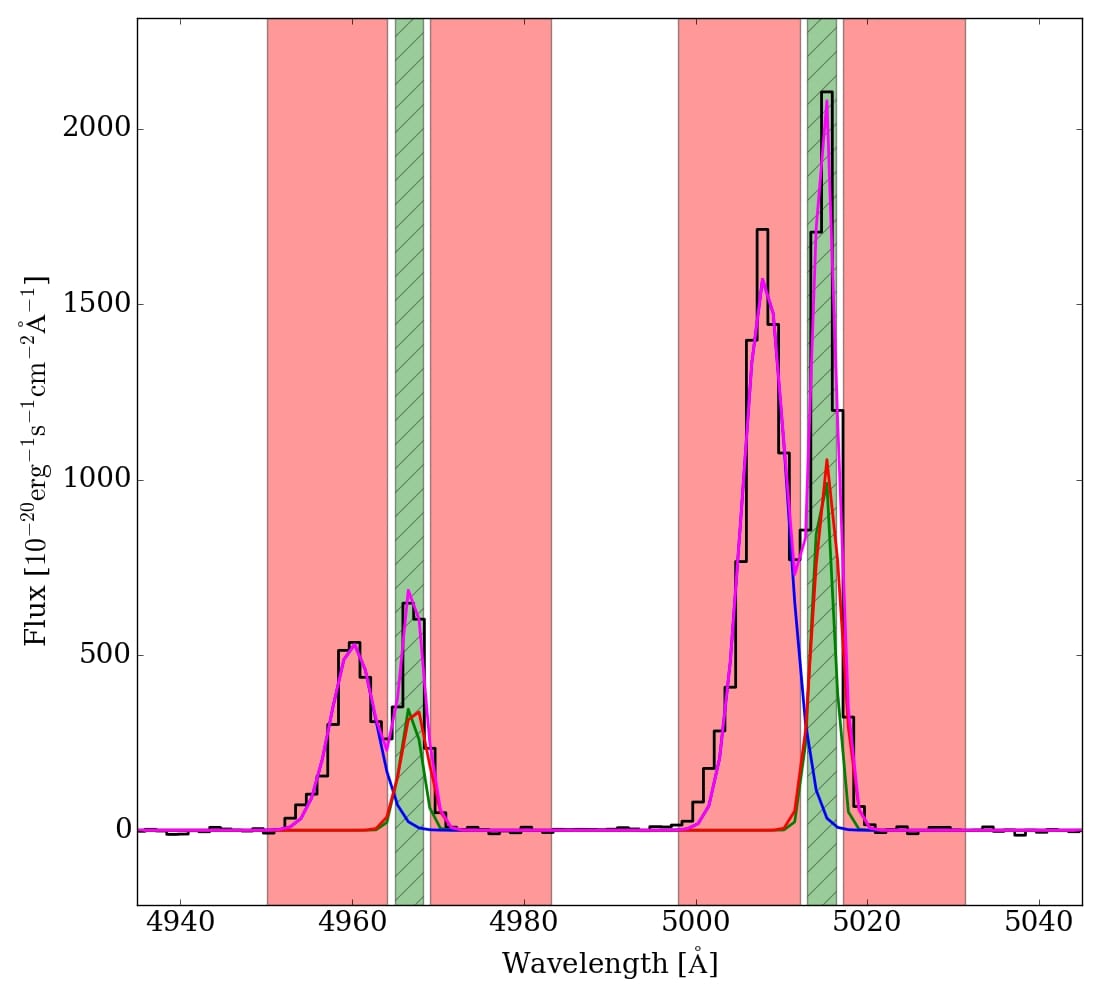}
    \caption{\oiii$\lambda\lambda$4959,5007 doublet in a single spaxel spectrum of the continuum-subtracted cube of the Circinus galaxy. The disc ($-100 \rm{~km/s} < \rm{v} < 100 \rm{~km/s}$) and outflow ($-1000 \rm{~km/s} < \rm{v} < -150 \rm{~km/s}$ and $150 \rm{~km/s} < \rm{v} < 1000 \rm{~km/s}$) components taken into account, highlighted in green with stripes and red, respectively, are superimposed to show an example of our procedure. The disc component is centred to the value of the stellar velocity in this spaxel ($\sim460\rm{~km/s}$), obtained previously by fitting the original data cube with \ppxf. The magenta line represents the result of the three-component fit (blue, green, and red Gaussians), performed with \mpfit \,to reproduce the observed line profiles.}
    \label{fig:example}
   \end{figure}

Figure~\ref{fig:example} shows a single spaxel spectrum extracted from the Circinus galaxy datacube to exemplify our division between disc (in green with stripes) and outflow (in red) components for the \oiii$\lambda\lambda$4959,5007 doublet. In many works, disc and outflow are separated according to the width of the two Gaussian components used in fitting the lines (narrower and broader, respectively). In our analysis, this approach is not feasible since disc and outflow components are both present in many spaxels. Furthermore, in the central regions the line profiles are very complex, and their fit requires three or four Gaussians.  This means that the distinction between disc and outflow through the width of the fitted components only is not trivial. 
We note that our approach is not a rigorous method for discriminating between the gas in the outflow and the disc, but a convenient way to analyse in detail the outflow features, as we  show in the following. 
    
\begin{figure*}
\centering
    \includegraphics[width=2\columnwidth]{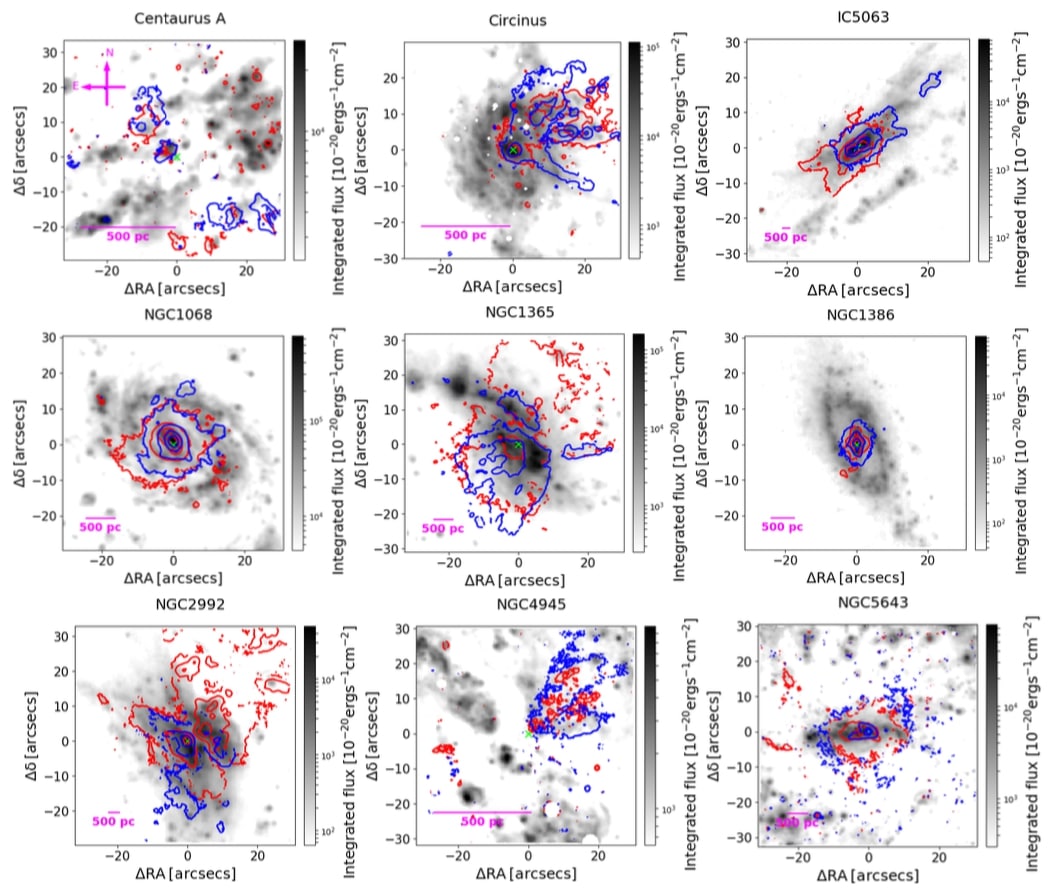}
    	\caption{H$\alpha$ disk component flux maps (not corrected for dust reddening), for all the galaxies - namely, Centaurus~A, Circinus, IC~5063, NGC~1068, NGC~1365, NGC~1386, NGC~2992, NGC 4945 and NGC~5643. \oiii$\lambda5007$ outflow component flux contours are superimposed for all the galaxies. Note that the blueshifted and redshifted outflow emission (in blue and red, respectively) is often extended in a kiloparsec-scale conical or biconical distribution. For each velocity bin, we show only the spaxels with a SNR~$>5$. The magenta bar represents a physical scale of $\sim500$~pc. East is to the left, as shown in the first image on the left. The white circular regions are masked foreground stars. The cross marks the position either of the Type 1 nucleus (i.e., peak of the broad H$\alpha$ emission), if present, or the peak of the continuum in the wavelength range $6800-7000$~\AA.}
        	\label{fig:vbin}
   \end{figure*}

Because the H$\alpha$ emission is in general dominated by the disc, while the outflow is enhanced in \oiii \, (e.g. \citealt{venturi2018}), in Fig.~\ref{fig:vbin} we show the H$\alpha$ disc component flux maps, superimposing the \oiii$\lambda5007$ outflow component flux contours (not corrected for dust reddening) for all the galaxies, using the method described above to discriminate between the two components. The blueshifted and redshifted outflowing components are indicated in blue and red, respectively.
For each velocity bin, we only select the spaxels with a S/N~$>5$, computed by dividing the integrated flux in the velocity channels by the corresponding noise. The noise is estimated from the standard deviation of the data-model residuals of the fit around each line (within a range about 60 to 110~\AA \,wide, depending on the line). Looking at Fig.~\ref{fig:vbin}, it can be seen that the disc flux maps and the outflow contours are clearly different from each other. The outflow component is often extended in a kpc-scale conical or biconical distribution, as  can be clearly seen in Circinus (north-west cone), IC~5063 (north-west and south-east cones), NGC~2992 (north-west and south-east cones), NGC~4945 (north-west and south-east cones), and NGC~5643 (east and west cones). In Centaurus~A the outflow component is mainly distributed in two cones (direction north-east and south-west) in the same direction of the extended double-sided jet revealed both in the radio and X-rays (e.g. \citealt{hardcastle2003}), and located perpendicularly with respect to the gas in the disc component. Unfortunately, since for this galaxy we cannot take the stellar velocity as a reference, in some regions we underestimate the gas velocity of the disc component. Consequently, a portion of the disc is still present in the outflowing component. Also in NGC~1365, the outflow component flux map has a biconical shape extended in the south-east and north-west directions, while the disc component appears to be completely different, being dominated by an elongated circumnuclear SF ring and by the bar. Unfortunately, similar to Centaurus~A, a portion of the disc is still present in our outflowing gas selection, because of the high velocity reached by the gas along the bar (see \citealt{venturi2018} for more details). NGC 1068 is almost face-on, allowing us to admire the spiral arms of the disc, traced by the disc component, and preventing the outflowing component from having a clear biconical distribution, but to be broadly extended in all the inner region of the observed FOV. As reported in Sect.~\ref{sec:sample1}, both the ionised and molecular outflow already observed in this galaxy are observed in the north-east and south-west directions (e.g. \citealt{cecil2000, burillo2014}).
Finally, in NGC~1386 the outflow component does not show a well-defined conical distribution, but appears to be located in the very inner region of the galaxy, with two elongated structures to the north and south of the nucleus, corresponding to nuclear bipolar outflows already revealed by \citet{ferruit2000} and \citet{lena2015}.

We note that in almost all galaxies the (bi)conical outflow is detected in both its blueshifted and redshifted components, which are often overlapping. For example, this can be clearly seen in NGC~4945, where the north-west cone has approaching velocities at its edges and receding ones around its axis, while the south-east cone has the opposite behaviour, with receding velocities at its edges and approaching ones around its axis. This was already revealed by \citet{venturi2017} through the kinematical maps of the \nii \, line emission.

\subsection{Extinction, density, and ionisation parameter}\label{sec:gas_properties}
\begin{figure}
   \centering
   \includegraphics[width=1.\columnwidth]{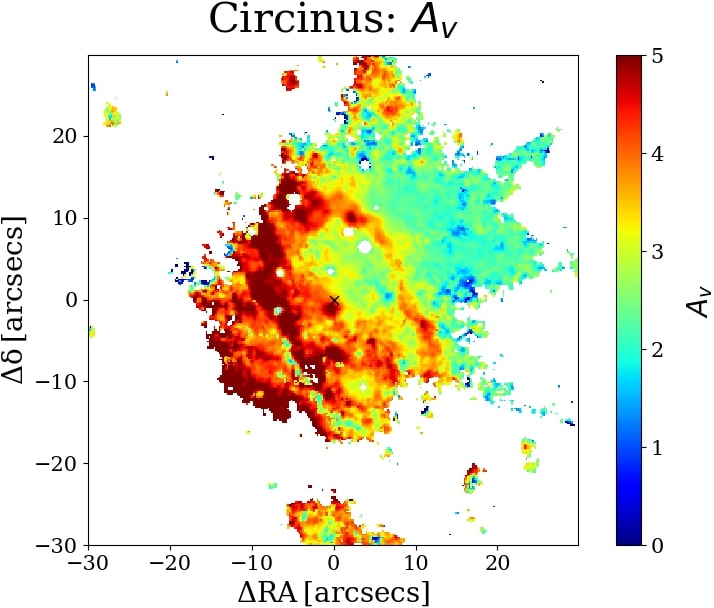}
   \centering
   \includegraphics[width=1.\columnwidth]{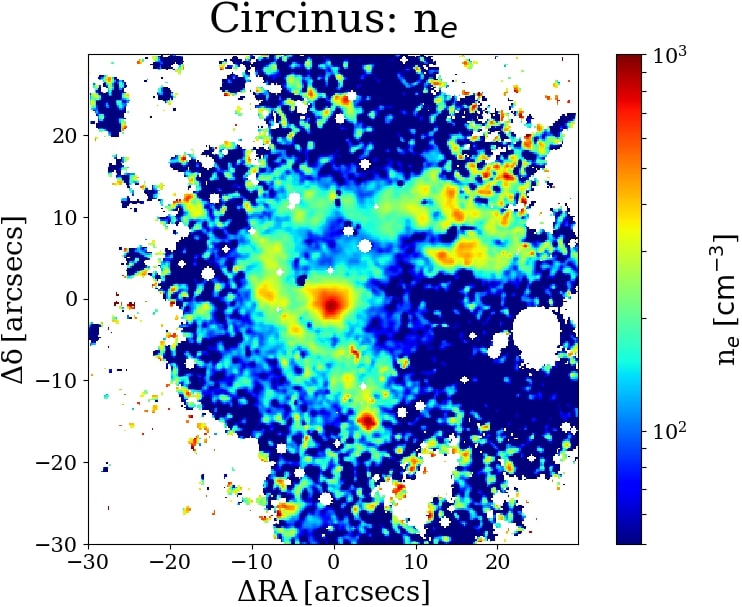}
    \centering
    \includegraphics[width=1.\columnwidth]{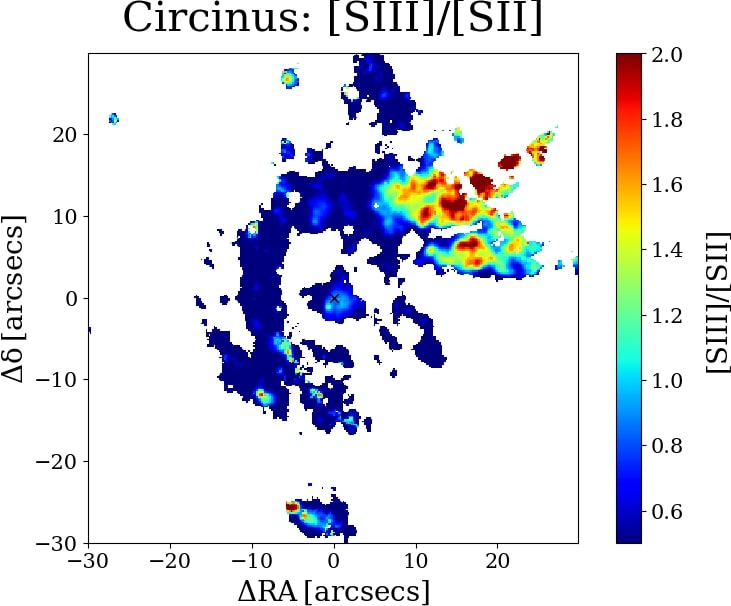}
        \caption{ISM properties in the Circinus galaxy. Top panel: Map of the total extinction in V band A$_{\rm V}$ obtained from the Balmer decrement H$\alpha$/H$\beta$. Only spaxels with H$\alpha$ and H$\beta$ S/N~$>5$ are shown. Central panel: Map of the total electron density measured from the \sii$\lambda$6717/\sii$\lambda$6731 ratio. Only spaxels with \sii$\lambda$6717 and \sii$\lambda$6731 S/N~$>5$ are shown. Bottom panel: Map of the \siii$\lambda\lambda$9069,9532/\sii$\lambda\lambda$6717,31 ratio, proxy for the ionisation parameter. Only spaxels with \siii$\lambda\lambda$9069,9532 and \sii$\lambda\lambda$6717,31 S/N~$>5$ are shown.}
    \label{fig:circinus_prop}
    \end{figure}

We calculated the visual extinction A$_{\rm V}$ through the Balmer decrement H$\alpha$/H$\beta$, assuming a \citet{calzetti2000} attenuation law, with $R_{\rm{V}}=3.1$ (i.e. galactic diffuse ISM) and a fixed temperature of $T_e = 10^4$~K. The top panel of Fig.~\ref{fig:circinus_prop} illustrates the extinction map of Circinus obtained for the total line profile (without separating disc and outflow component), where we exclude all the spaxels that have S/N~$<5$ on the emission lines involved. In this case, we defined the S/N of a line as the ratio between the peak value of the fitted line profile and the standard deviation of the data-model residuals of the fit around the line. The highest values A$_{\rm V}\gtrsim 5$ come from the galaxy disc, while the conical outflow is characterised by A$_{\rm V}\gtrsim 2$. 
The extinction maps for the other galaxies are shown in Fig.~\ref{afig:av_maps}, showing that the highest values of A$_{\rm V}$ come mainly from dust lanes (e.g. Centaurus~A and NGC~2992), gas discs as shown for Circinus (e.g. NGC 4945), spiral arms (e.g. NGC~1068 and NGC~1386), or gas flowing along the bar (e.g. NGC~1365).

\begin{figure}
   \centering
   \includegraphics[width=1.05\columnwidth]{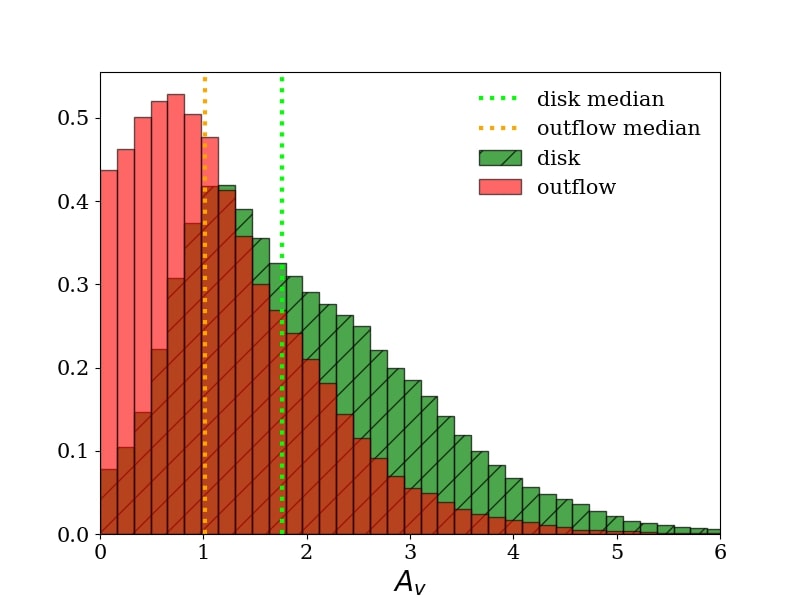}
   \centering
  \includegraphics[width=1.05\columnwidth]{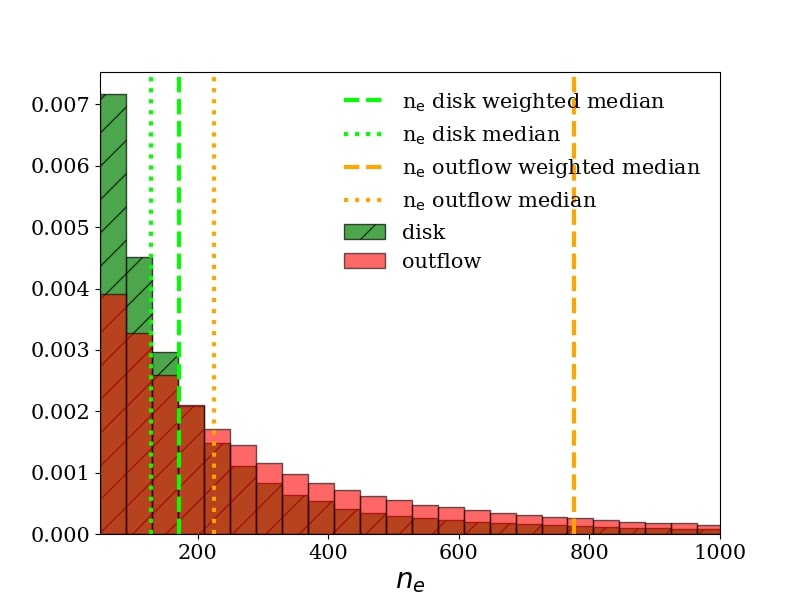}
    \centering
    \includegraphics[width=1.05\columnwidth]{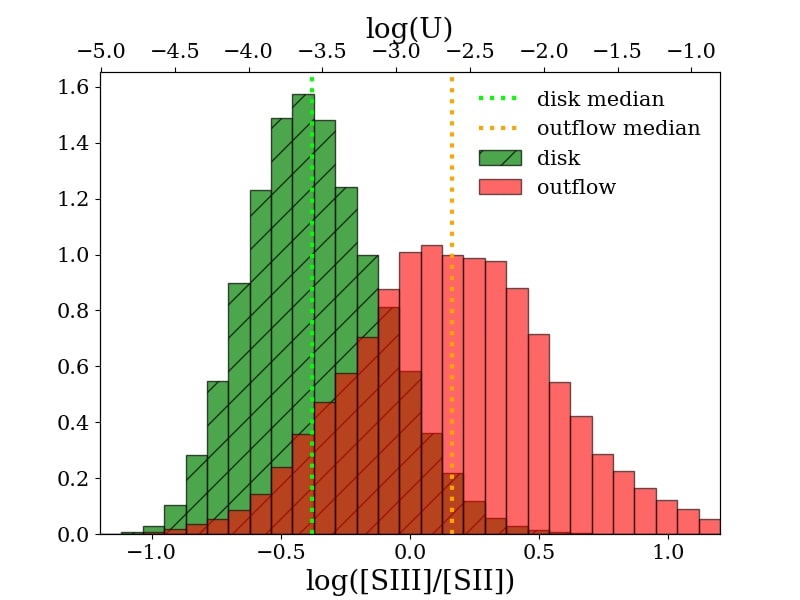}
       \caption{Comparison of the disc (green with stripes) and outflow (red) component distributions for all the MAGNUM galaxies in terms of visual extinction A$_{\rm V}$, density $n_{\rm e}$, and log(\siii/\sii) \,(top, middle, and bottom panel, respectively). The two distributions are normalised such that the integral over the range is 1. For each velocity bin, we take into account only the spaxels with a S/N~$>5$ for the emission lines involved. The light green and orange dotted lines represent the median of the disc and outflow distributions, respectively. The dashed lines in the central panel represent the median value weighted on the \sii \, line flux. The top x-axis of the bottom panel translates the \siii/\sii \,ratio into log($U$), following the relation provided by \citet{diaz2000}.}
    \label{fig:histograms}
    \end{figure}

In order to analyse in detail these differences, we measured two values of A$_{\rm V}$ in each spaxel for every galaxy, one for the disc and one for the outflow component. To do so, for each velocity bin we selected only the spaxels with a S/N~$>5$, and we discarded all the velocity bins where H$\alpha$/H$\beta < 2.86$ (the theoretical value associated with A$_{\rm V} = 0)$.
The top panel of Fig.~\ref{fig:histograms} shows the flux spatial distributions of A$_{\rm V}$ for the disc (in green with stripes) and outflow (in red) components of all the MAGNUM galaxies, with the median of the two distributions superimposed (dotted line). Interestingly, the two distributions are shifted in A$_{\rm V}$ relative to each other, with higher values of extinction in the disc (median: A$_{\rm V}\sim 1.75$ - H$\alpha$/H$\beta\sim5.51$) and lower in the outflow (median: A$_{\rm V} \sim 1.02$ - H$\alpha$/H$\beta\sim4.16$). Therefore, the majority of the outflowing gas appears to be less affected by dust extinction than the disc. We have verified that there is no large distinction between the redshifted and blueshifted outflowing gas distributions due to projection effects, probably due to the fact that redshifted and blueshifted velocities coexist in each cone of the outflow because of geometrical effects (Venturi et al. in preparation). 

In order to estimate the gas density, we make use of the optical \sii$\lambda$6717/\sii$\lambda$6731 ratio, converting it to an electron density, using the \citet{osterbrock2006} model, assuming a temperature of $T_e =10^4$~K. This tracer is density-sensitive in the range $50$~cm$^{-3} \lesssim n_{\rm e} \lesssim 5000$~cm$^{-3}$, which falls within the typically estimated range of densities in the extended NLR (e.g. \citealt{perna2017}). Below and above these densities, the flux ratio of the doublet saturates. The central panel of Fig.~\ref{fig:circinus_prop} reports the density map of Circinus, obtained for the total line profile, that appears clearly non-uniform, as found by \citet{kakkad2018}, with enhanced values of density up to $n_{\rm e} \sim 10^3$~cm$^{-3}$ located in the inner regions of the galaxy and along the north-western outflowing cone. The density maps of the other galaxies in the sample are reported in Fig.~\ref{afig:ne_maps}. The majority of them shows the same characteristics found in Circinus. The highest densities, with peaks of up to $n_{\rm e} \sim 10^3$~cm$^{-3}$ (e.g. IC~5063, NGC~1068, NGC~1386), are mainly found in the central regions of the galaxies. High densities ($10^2$~cm$^{-3} <n_{\rm e} < 10^3$~cm$^{-3}$) are also found along the outflowing cone axis, as  is clearly visible in the north-east and south-west directions in NGC~1068 and in the double-side cone of NGC 4945 (north-west--south-east direction). On the contrary, NGC~1365 has enhanced values of density around the circumnuclear star-forming ring (see also \citealt{kakkad2018, venturi2018}), consistent with recent results showing a correlation between density and the location of the SF regions (e.g. \citealt{westmoquette2011, westmoquette2013, mcleod2015}). 
The central panel of Fig.~\ref{fig:histograms} shows the density distribution of all the galaxies for the disc (in green with stripes) and outflow (in red) components, separately. The two distributions peak at the minimum value of the electron density range taken into account (i.e. $n_{\rm e} = 50$~cm$^{-3}$), but the disc median value is $n_{\rm e} \sim 130$~cm$^{-3}$ against $n_{\rm e} \sim250$~cm$^{-3}$ for the outflowing gas, and thus the outflow is generally denser than the disc gas in MAGNUM galaxies. 

Previous works (e.g. \citealt{holt2011,arribas2014,villarmartin2014,villarmartin2015}) have found  very high reddening and densities associated with ionised outflows in local objects (e.g. H$\alpha$/H$\beta \sim 4.91$ and $n_{\rm e} \gtrsim 1000$~cm$^{-3}$, \citealt{villarmartin2014}). 
Concerning the reddening, although we find that the outflowing gas is generally less affected by dust extinction than the disc, the median value of the total distribution is significantly affected by dust (H$\alpha$/H$\beta \sim4.16$), with tails up to H$\alpha$/H$\beta \gtrsim 6$. Similarly, the outflow density of MAGNUM galaxies is higher than the values in the disc gas, but appears to be far lower than the values found by these authors. This could stem from the fact that the galaxies studied by \citealt{holt2011,arribas2014} are local luminous or ultra-luminous infrared galaxies (U/LIRGs), and those of \citealt{villarmartin2014,villarmartin2015} are highly obscured Seyfert 2, thus sampling sources that are more gas and dust rich compared to our sample.
However, our values are also lower  than the outflow densities found in \citet{perna2017} ($n_{\rm e} \sim 1200$~cm$^{-3}$), who targeted optically selected AGNs from the SDSS, and \citet{forsterschreiber2018}, who presented a census of ionised gas outflows in high-z AGN with the KMOS$^{\rm 3D}$ survey ($n_{\rm e} \sim 1000$~cm$^{-3}$). A possible explanation could be related to the high quality of our MUSE data, which also allows us to detect  the faint \sii \, emission associated with lower density regions. If we calculate the median densities of the disc and outflow components, weighting for the \sii \, line flux, we obtain higher values ($n_e \sim 170$~cm$^{-3}$ and $n_e \sim 815$~cm$^{-3}$, for disc and outflow, respectively). This shows that previous outflow density values from the literature could be biased towards higher $n_e$ because they are based only on the most luminous outflowing regions, characterised by a higher S/N. This could also mean that outflows at high-z could be far more extended than the values we observe.

Finally, we traced the ionisation parameter $U$, defined as the number of ionising photons $S_{*}$ per hydrogen atom density $n_H$, divided by the speed of light $c$, making use of the \siii$\lambda\lambda$9069,9532/\sii$\lambda\lambda$6717,31 ratio\footnote{Both emission lines were corrected for dust attenuation prior to deriving the ionisation parameter.} (e.g. \citealt{diaz2000}). Since  \siii$\lambda$9532 is not covered by the wavelength range observed by MUSE, we adopted a theoretical ratio of \siii$\lambda\lambda$9532/\siii$\lambda\lambda$9069 = 2.5 \citep{vilchez1996}, fixed by atomic physics. The parameter $U$ represents a measure of the intensity of the radiation field, relative to gas density. It can be traced using the ratios of emission lines of different ionisation stages of the same element: the larger the difference in ionisation potentials of the two stages, the better the ratio will constrain $U$. 
The \siii$\lambda\lambda$9069,9532/\sii$\lambda\lambda$6717,31 ratio is a more reliable diagnostic than \oiii$\lambda\lambda$4959,5007/\oii$\lambda\lambda3726,28$, since it is largely independent on metallicity \citep{kewley2002}. The bottom panel of Fig.~\ref{fig:circinus_prop} shows the \siii/\sii \, ratio map of Circinus where  \siii/\sii \,reaches values as high as $\sim 2$ along the conical outflow. Fig.~\ref{afig:ionu_maps} reports the  \siii/\sii \,maps for the other galaxies. 
Similar to Circinus, several other galaxies in our sample (NGC~1068, IC~5063, NGC~1386, NGC~2992, and NGC~5643) also show higher values of  \siii/\sii \,along the outflow axis. In addition, \siii/\sii \,is high in the star-forming region embedded in the gas disc in Centaurus~A, and located in isolated blobs in NGC~1365. Unfortunately, the \siii/\sii \,ratio is poorly constrained in NGC~4945.  
The bottom panel of Fig.~\ref{fig:histograms} shows a comparison between the distribution of \siii/\sii \,associated with the disc component (in green with stripes) and with the outflow (in red). The top x-axis translates the \siii/\sii \,ratio into log($U$), following the relation provided by \citet{diaz2000}, assuming that the gas is optically thin \citep{morisset2016}. 
In conclusion,  log(\siii/\sii) \,is observed in the range from $\sim-1$ to $\sim1$, with higher values associated with the outflowing gas (median: log(\siii/\sii)~$\sim0.16$, log($U$)~$\sim-2.75$) with respect to the disc (median: log(\siii/\sii)~$\sim-0.38$, log($U$)~$\sim-3.60$), where it  traces mainly star-forming regions. The median value found for the disc component is slightly lower than the ionisation parameters found in nearby star-forming galaxies in other studies (e.g. \citealt{dopita2006, liang2006, kewley2008, nakajima2014, bian2016}), obtained through the \oiii$\lambda\lambda$4959,5007/\oii$\lambda\lambda3726,28$. 
We also calculated  the median values of the extinction and ionisation parameter distributions, weighting for the line fluxes of the involved lines, finding negligible ($\Delta A_{\rm V} \sim -0.02$, $\Delta \rm{log(\siii/\sii)} \sim +0.03$) and small differences ($\Delta A_{\rm V} \sim - 0.20$, $\Delta \rm{log(\siii/\sii)} \sim + 0.25$) with respect to the median values shown in Fig.~\ref{fig:histograms}, for the disc and outflow component, respectively.

In summary, we find that the disc and outflow are characterised by different properties. In the next section, we  investigate the ionisation properties of the gas separately for the two kinematic components.
    
\subsection{Spatially and kinematically resolved BPT}\label{sec:BPTs}
Spatially and kinematically resolved BPT diagrams allow us to explore the dominant contribution to ionisation in each spaxel in the disc and outflow components separately (a similar approach has been used by e.g. \citealt{westmoquette2012,mcelroy2015,karouzos2016b}). 

The left and middle panels of Fig.~\ref{fig:cir_bpt} show the \nii- and \sii-BPT diagrams of Circinus for the disc and outflow components, respectively. For each velocity bin, we select only the spaxels with a S/N~$>5$, computed by dividing the integrated flux in the velocity channels for each emission line by the corresponding noise. The dashed curve is the boundary between star-forming galaxies and AGN defined by \citet{kauffmann2003}, while the solid curve is the theoretical upper limit on SF line ratios found by \citet{kewley2001}. The dotted line, instead, is the boundary between Seyferts and LI(N)ERs introduced by \citet{kewley2006}. The dominant source of ionisation is colour-coded: shades of blue for SF,  green for intermediate regions in the \nii-BPT and LI(N)ER in the \sii-BPT, and  red for AGN-like ionising spectra, as a function of the x-axis line ratios (darker shades means higher x-axis line ratios). The LI(N)ER-like excitation can be  due either to shock excitation (e.g. \citealt{dopita1995}) or to hard-X radiation coming from AGN and to hot evolved (post-asymptotic giant branch) stars (e.g. \citealt{singh2013,belfiore2016a}). The corresponding position on the map of the outflowing gas component, colour-coloured  based on the different sources of ionisation, is shown in the right panels of Fig.~\ref{fig:cir_bpt}. In the background of all the pictures (black dots in the BPTs and shaded grey in the corresponding maps), we show the disc and outflow components together to allow a better visual comparison.

Looking at Fig.~\ref{fig:cir_bpt}, we note that the outflow spans a wider range of \nii/H$\alpha$ and \sii/H$\alpha$, including lower and higher values compared to the disc. Specifically, the highest and lowest values of low-ionisation line ratios (LILrs; i.e. \nii/H$\alpha$ and \sii/H$\alpha$), displayed in dark red and orange, are prominent in the AGN/LI(N)ER-dominated outflow component, while they are not observable in the disc component.
This can also be seen  looking at the \nii- and \sii-BPT diagrams of the other galaxies, shown in Figs.~\ref{afig:bptnii_gal_1}, \ref{afig:bptnii_gal_2}, and Fig~\ref{afig:bptsii_gal_1}, \ref{afig:bptsii_gal_2}, respectively\footnote{\nii- and \sii-BPT diagrams in the Appendix were made taking into account the smoothed datacubes. An exception is made for NGC~1365, where the biconical outflow can be well traced only in the Voronoi-binned cube.}.
In the following we analyse the regions showing the lowest  (Sect.~\ref{sec:lLILr}) and the highest  (Sect.~\ref{sec:hLILr}) LILrs.

\subsubsection{Outflowing gas: the lowest \nii/H$\alpha$-\sii/H$\alpha$ line ratios}\label{sec:lLILr}
As Fig.~\ref{fig:cir_bpt} shows, the lowest \nii/H$\alpha$ and \sii/H$\alpha$ line ratios (in orange, with log(\nii/H$\alpha$)~$\lesssim -0.5$ and log(\sii/H$\alpha$)~$\lesssim -0.8$) in the AGN ionisation regime belong almost exclusively to the outflowing gas.
Looking at the right panel of Fig.~\ref{fig:cir_bpt}, it can be seen that these values correspond to the innermost parts of the northwestern cone (in orange)\footnote{Unfortunately, part of the outflowing gas in the west cone of Circinus is oriented along the plane of the sky, and thus has low line-of-sight velocity. Consequently, the disc component is partly contaminated by the outflow, explaining the presence of the orange dots in the left panels of the BPT diagrams in Fig.~\ref{fig:cir_bpt}.}.

These low values of \nii/H$\alpha$ and \sii/H$\alpha$ (with logarithmic values down to $\sim -1$) in the AGN-dominated region are not  restricted to the Circinus outflowing gas, but appear to be typical outflow signatures in almost all the galaxies of our sample (NGC~2992, NGC~1365, IC~5063, NGC~1068), and,  albeit less clearly, in Centaurus~A, NGC~1386, and NGC~5643 (see  left panels of Figs.~\ref{afig:bptnii_gal_1} and \ref{afig:bptnii_gal_2}, for \nii-BPT diagrams, Figs.~\ref{afig:bptsii_gal_1} and \ref{afig:bptsii_gal_2}, for \sii-BPT diagrams).
Specifically, the lowest LILrs dominate the majority of the visible north-west cone of NGC~2992, which extends out to $\sim 5$~kpc, and the innermost parts of the cones in most of the other galaxies (north-west cone in NGC~1365, north-west and south-east directions in IC~5063, north-east and south-west directions in NGC~1068, east and west directions in NGC~5643), highlighted in orange in the right panels of Figs.~\ref{afig:bptnii_gal_1} and \ref{afig:bptnii_gal_2}, Figs.~\ref{afig:bptsii_gal_1} and \ref{afig:bptsii_gal_2}, for the \nii- and the \sii-BPT diagrams, respectively.
On the contrary, these features seem to be completely absent in NGC 4945. 

\subsubsection{Outflowing gas: the highest \nii/H$\alpha$ and \sii/H$\alpha$ line ratios}\label{sec:hLILr}
\begin{figure*}
    \begin{minipage}{0.66\columnwidth}
                \includegraphics[width=1.\columnwidth]{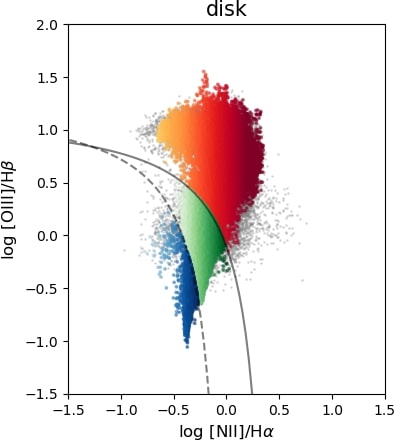}
    \end{minipage} 
        \begin{minipage}{0.66\columnwidth}
                \includegraphics[width=1.\columnwidth]{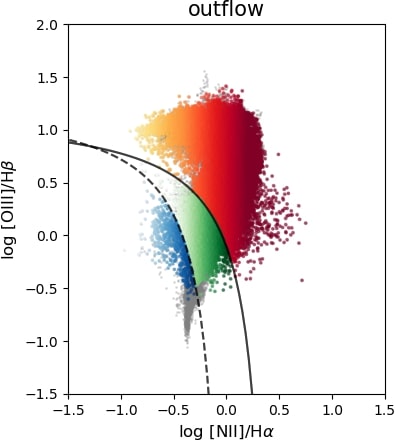}
    \end{minipage} 
        \begin{minipage}{0.66\columnwidth}
         \includegraphics[width=1.\columnwidth]{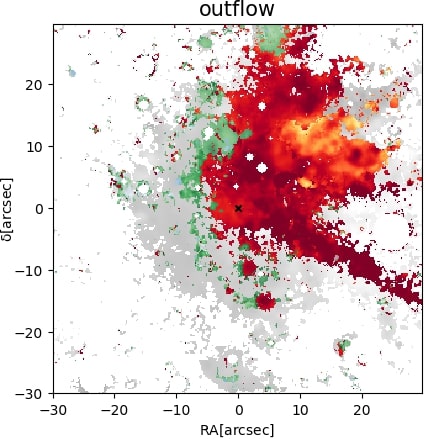}
   \end{minipage}      
       \begin{minipage}{0.66\columnwidth}
                \includegraphics[width=1.\columnwidth]{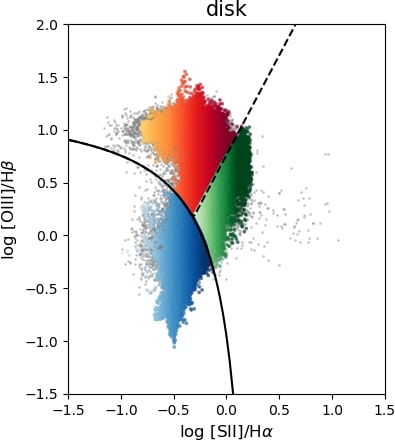}
    \end{minipage}
           \begin{minipage}{0.66\columnwidth}
                \includegraphics[width=1.\columnwidth]{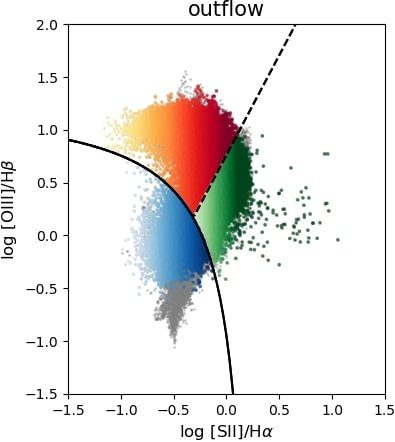}
    \end{minipage}
        \begin{minipage}{0.66\columnwidth}
         \includegraphics[width=1.\columnwidth]{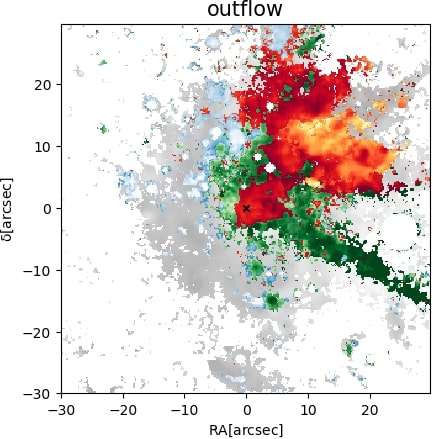}
   \end{minipage}    
         \caption{Left and central panels: \nii- and \sii-BPT diagrams for the disc and outflow components, on the left and on the right respectively, of Circinus. Blue denotes SF-dominated regions, green intermediate regions in the \nii-BPT and LI(N)ER in the \sii-BPT, and red AGN-like ionised spectra, colours-coded  as a function of the x-axis line ratios (darker shades means higher x-axis line ratios). The black dashed curve is the boundary between star-forming galaxies and AGN defined by \citet{kauffmann2003}, while the black solid curve is the theoretical upper limit on SF line ratios found by \citet{kewley2001}. The black dotted line is the \citet{kewley2006} boundary between Seyferts and LI(N)ERs. Right panels: \nii- and \sii-BPT maps of the outflowing gas component, colour-coded according to BPTs. In the background of all the pictures (grey dots in the BPTs and shaded grey in the corresponding maps), we show the disc and outflow component together. For each velocity bin, we select only the spaxels with a S/N~$>5$ for all the lines involved.}
                \label{fig:cir_bpt}
   \end{figure*}

Looking at Fig.~\ref{fig:cir_bpt}, the highest \nii/H$\alpha$ and \sii/H$\alpha$ line ratios (in darker red and green, respectively), with values higher than 0, appear to come from the edges of the outflowing cone.

As noted above for the other ratios, this feature is not only typical of Circinus, but appears to characterise almost all of the sample (NGC 4945, NGC~2992, NGC~1068, Centaurus~A, IC~5063, NGC~1386 and NGC~5643), as  can be seen in the left panels of Figs.~\ref{afig:bptnii_gal_1} and \ref{afig:bptnii_gal_2}, Figs.~\ref{afig:bptsii_gal_1} and \ref{afig:bptsii_gal_2}, for \nii- and \sii-BPT diagrams, respectively. 
For instance, it is clearly visible in NGC 4945, where -- as in Circinus -- the high LILrs  come from the external edges of both the north-western and the south-eastern outflowing cones, even though in the latter only the edges are barely observable because of the low H$\beta$ S/N. 
In NGC~1068, IC~5063 and in NGC~5643, this enhancement is extended from the edges of the cones to a region almost perpendicular to the outflow axis\footnote{In NGC~1068, IC~5063 and NGC~5643 the outflow axis is in the direction north-east and south-west, north-west and south-east, and east and west, respectively.} where a conical dark red region (AGN-dominated) in \nii-BPT and a conical dark green region (LI(N)ER-dominated) in \sii-BPT can be distinguished.

On the contrary, it can be seen that in NGC~1365 the highest LILr stem from gas located in the disc component in the vicinity of the outflowing cone edges. However, we believe that the distinction between disc and outflow components is not completely defined in this galaxy. Part of the outflowing gas could have an apparent low line-of-sight velocity because it is  oriented along the plane of sky, contaminating the disc component. 

\section{Discussion}\label{sec:discussion}
In the analysis of the kinematically and spatially resolved BPT diagrams for our sample of galaxies (see Sect. \ref{sec:BPTs}), we identify two main features in the outflowing gas:  a decrease and an enhancement in the \nii/H$\alpha$ and \sii/H$\alpha$ line ratios, coming from distinct regions of the outflowing cones.
In the following, we discuss possible explanations for these features, such as a combination of optically thin and optically thick clouds or shocks and/or a hard filtered radiation field from the AGN. To do this, we compare ionisation state and gas velocity dispersion with \nii- and \sii- BPT diagrams comprising the whole MAGNUM sample; the only  exception is NGC~1068,  which  shows a different behaviour with respect to the others concerning the \siii/\sii \, ratio (see Figs.~\ref{afig:bptnii_ionu_gal} and \ref{afig:bptsii_ionu_gal}) and has a far higher velocity dispersion (see Figs.~\ref{afig:bptnii_sig_gal} and \ref{afig:bptsii_sig_gal}). Therefore, it will be discussed in a forthcoming paper.

\subsection{Possible scenarios for the lowest LILr}
\begin{figure*}
     \centering
     \begin{minipage}{2\columnwidth}
                \includegraphics[width=1.\columnwidth]{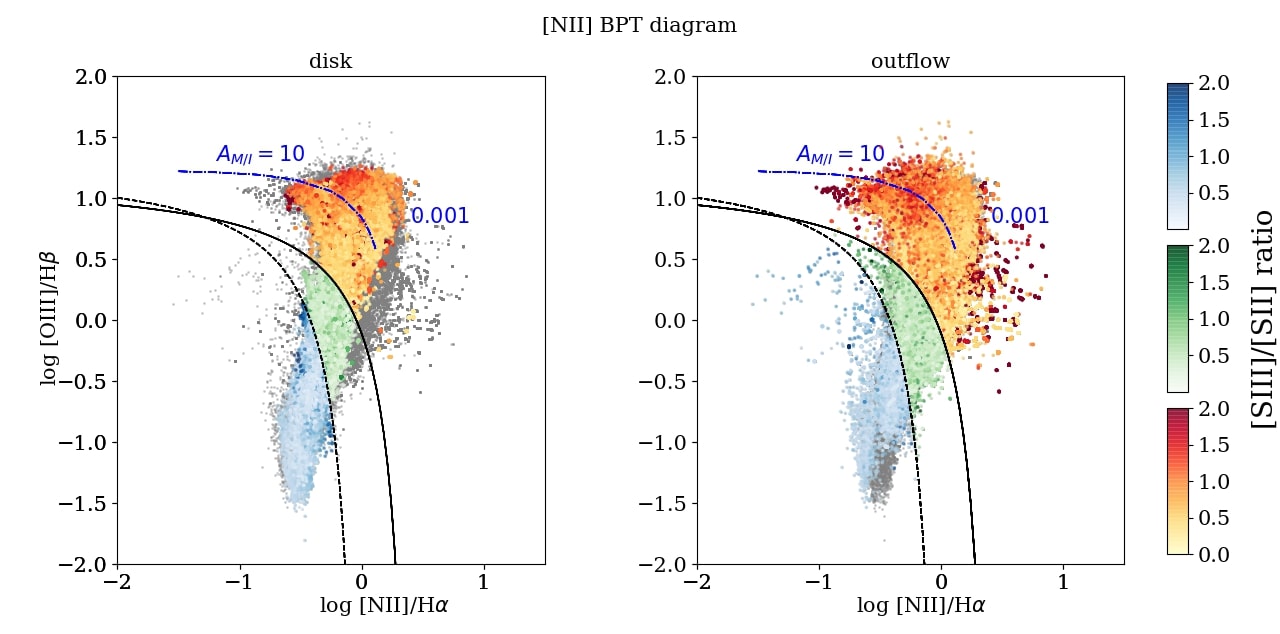}
      \end{minipage}
      \begin{minipage}{2\columnwidth}
                \includegraphics[width=1.\columnwidth]{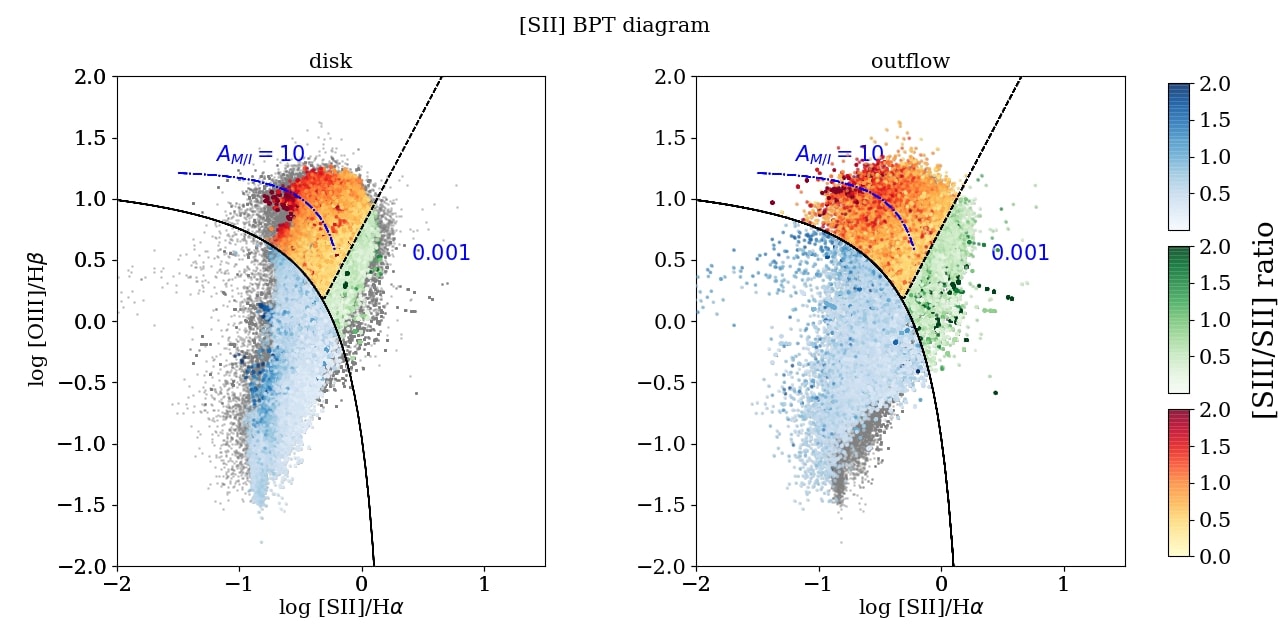}
     \end{minipage}
         \caption{\nii- and \sii-BPT diagrams for the disc and outflow components, on the left and the right respectively, of all the MAGNUM galaxies, apart from NGC~1068, colour-coded as follows: shades of blue for SF,  green for intermediate regions in the \nii-BPT and LI(N)ER in the \sii-BPT, and  red for AGN-like ionising spectra, as a function of the \siii/\sii \, line ratio (darker shades means higher \siii/\sii \, values).  
The black dashed curve is the boundary between star-forming galaxies and AGN defined by \citet{kauffmann2003}, while the black solid line is the theoretical upper limit on SF line ratios found by \citet{kewley2001}. The dash-dotted blue curve represents the Binette  A$_{M/I}$-sequence, dominated by MB clouds in the upper left region and by IB clouds going towards higher  LILrs. The black dotted line is the \citet{kewley2006} boundary between Seyferts and LI(N)ERs. The black dots in the BPTs are all the spaxels taken into account (both of the disc and of the outflow component) with a S/N~$>3$ for the line fluxes involved.}
   \label{fig:bpt_total_mod1}
   \end{figure*}

Figure~\ref{fig:bpt_total_mod1} shows the \nii- and \sii- BPT diagrams of all the MAGNUM galaxies (excluding NGC~1068), colour-coded as a function of the \siii$\lambda\lambda$9069,9532/\sii$\lambda\lambda$6717,31 ratio, which is a proxy for the ionisation parameter, as reported in the corresponding colour bars (darker shades means higher ionisation). Here, we use the Voronoi-binned cubes to increase the S/N and, for each Voronoi bin, we select only the velocity bins where the \siii \, and \sii \, line fluxes have a S/N~$>3$. 
The lowest LILrs (log(\nii/H$\alpha$)~$ \sim -1$ and log(\sii/H$\alpha$)~$\sim -1$), which  mainly trace the innermost regions of the outflowing gas, also correspond to the highest \siii/\sii \, line ratio (\siii/\sii~$\sim 2$), which decreases to values \siii/\sii~$\lesssim 0.5$ going towards higher LILr values. 

This observed variation in the degree of excitation can be interpreted as being due to diverse proportions of two populations of line emitting clouds, characterised by  different levels of opacity to the radiation, a matter-bounded (MB) and an ionisation-bounded (IB) component, first discussed by \citet{binette1996}. The former is only responsible for the high-ionisation lines (e.g. \oiii $\lambda\lambda$4959,5007), while the latter is also characterised by low-ionisation ones (e.g. \nii$\lambda\lambda$6548,84 and \sii$\lambda\lambda$6717,31). Specifically, a cloud is said to be MB if the outer limit to the H$^+$ region is marked by the outer edge of the cloud, which means that it is ionised throughout and is optically thin to the incident radiation field. On the other hand, the cloud is IB (or radiation bounded) if the outer limit to the H$^+$ region is defined by a hydrogen ionisation front, so both warm ionised and cold neutral regions coexist, and thus the H$^+$ region is optically thick to the hydrogen-ionising radiation, absorbing nearly all of it.
The original aspect of the \citet{binette1996} approach is the assumption that the IB clouds are photoionised only by hard photons which have leaked through the MB clouds, meaning that the input ionising spectrum impinging on the IB clouds has been partially filtered (see Fig.~3, \citealt{binette1996}). Hard photons interact much less effectively with the gas, being their energy far above the H$^0$ ionisation threshold. This means that, unlike the MB clouds which share the same excitation (i.e. the same ionisation parameter) being coupled with the radiation field, the IB clouds are not necessarily characterised by a unique value of ionisation parameter. Their modelling is parametrised by the ratio of the solid angles subtended by MB and IB clouds ($A_{M/I}$), which introduces a new free parameter:  the geometry of the system, which is often difficult to constrain. 

In Fig.~\ref{fig:bpt_total_mod1}, the blue dash-dotted line represents the Binette $A_{M/I}$ sequence \citep{binette1996,allen1999}, that closely matches  the differences in the line ratios that we observe: in the upper left region (A$_{M/I} \sim 10$), the MB component dominates, meaning that the gas is more optically thin, and is thus characterised by high-ionisation lines, such as \oiii \, and \siii, tracing the inner gas within the outflowing cones (see Sect.~\ref{sec:BPTs}); going to higher  LILrs, the optically thick IB clouds, characterised by a much lower excitation, start playing a major role ($A_{M/I} \sim 0.001$) both in the outflowing gas and in the disc gas. However, even though the MB-IB dichotomy accounts for the low LILr, it cannot explain the highest LILr that we observe in our sample.

Baldwin-Phillips-Terlevich diagrams colour-coded as a function of \siii/\sii \, for each galaxy can be seen in Fig.~\ref{afig:bptnii_ionu_gal} and Fig.~\ref{afig:bptsii_ionu_gal}, for the \nii- and \sii-BPT diagrams, respectively\footnote{The \nii- and \sii-BPT diagrams  given in the Appendix, apart from  NGC~1365, are made taking into account the smoothed datacubes, and not the Voronoi-binned ones that we use in this section to compare all the galaxies.}. The trend found between the \siii/\sii \, line ratio and low LILr is mainly visible in Circinus, NGC~2992, and NGC~5643, but also in NGC~1386, even though less clearly because of the cut in S/N. We note that NGC 4945, which is the only galaxy that lacks the MB clouds features, is characterised either by a very strong obscuration of AGN UV radiation or by an AGN that is highly deficient in its UV with respect to X-ray emission \citep{marconi2000}. This could explain why we see only the IB cloud component, ionised by hard photons filtered by the dust. 

\begin{figure*}
         \centering
           \begin{minipage}{2.\columnwidth}
           \centering
                \includegraphics[width=1.\columnwidth]{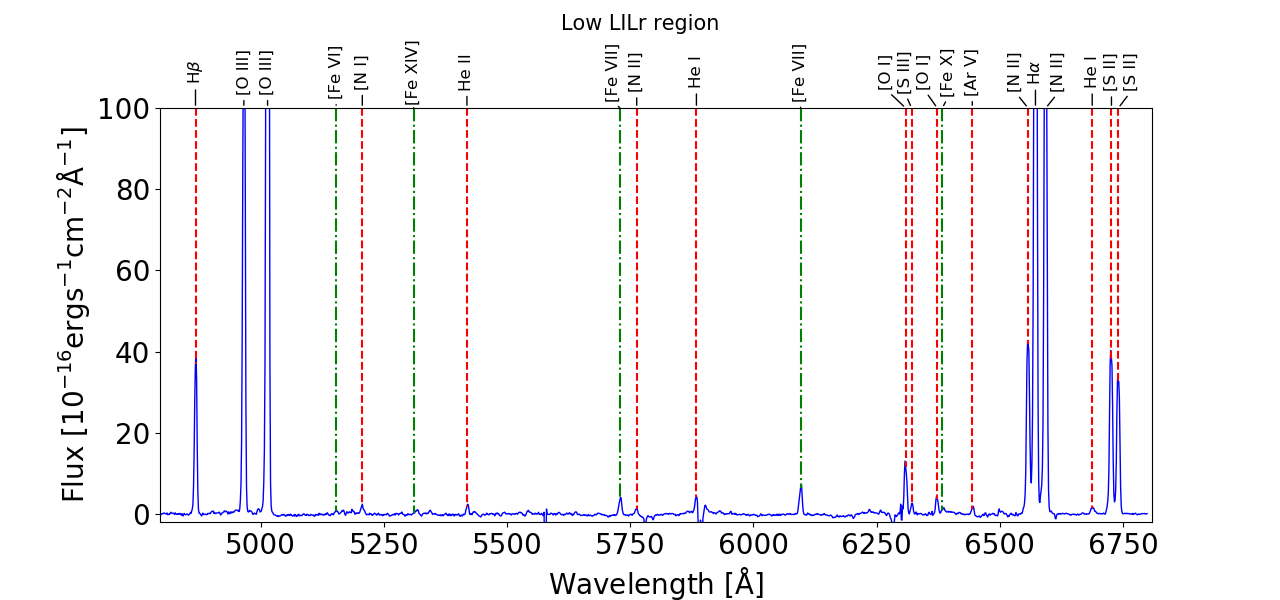}
        \end{minipage}
        \begin{minipage}{2.\columnwidth}
        \centering
                \includegraphics[width=1.\columnwidth]{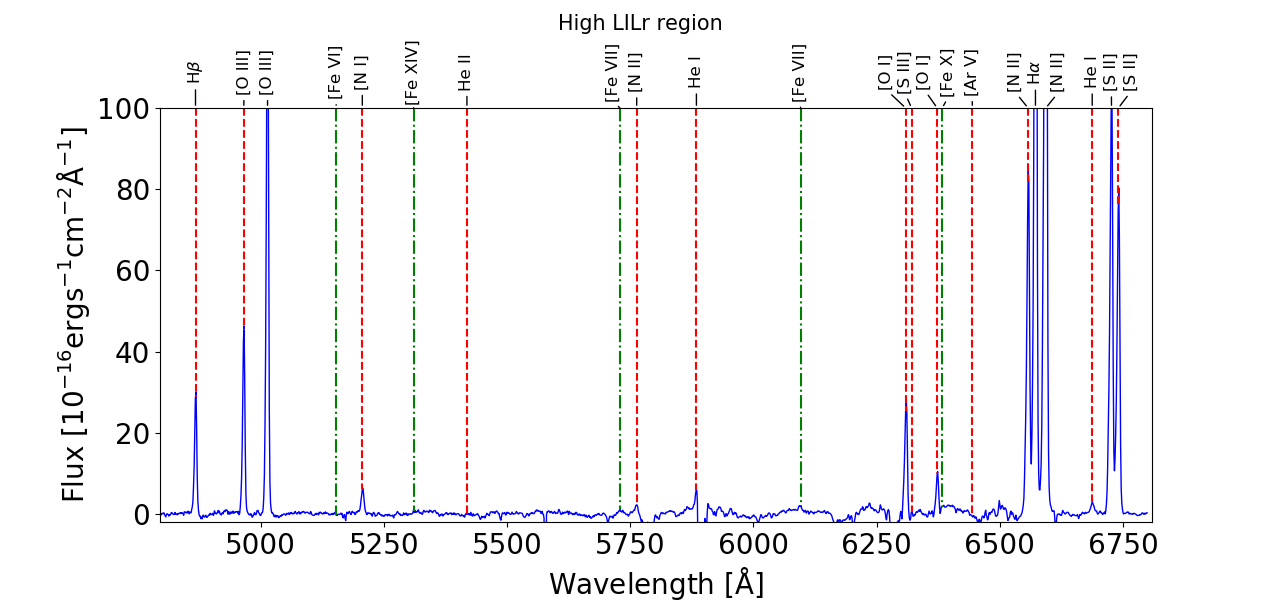}
        \end{minipage}
         \caption{Top panel: Spectrum of the region characterised by log(\nii/H$\alpha$)$<-0.2$ and log(\sii/H$\alpha$)$<-0.4$ in Circinus, corresponding to the innermost regions of the outflowing cone. Bottom panel: Spectrum of the region characterised by log(\nii/H$\alpha$)$>0.1$ and log(\sii/H$\alpha$)$>0$ in Circinus, mainly tracing the edges of the north-western outflowing cone. High-ionisation coronal lines (in green) are observed almost exclusively in the low LILr region, supporting a scenario where the inner parts of the outflowing gas in this galaxy are in optically thin highly ionised regions.}
         \label{fig:spectra}
   \end{figure*}

In  the top panel of Fig.~\ref{fig:spectra}, we show the stacked spectrum in the wavelength range 4850--7500 \AA \, of the region corresponding to the lowest LILr in Circinus. Specifically, we selected and summed  the AGN- and LI(N)ER-dominated spaxels from the
original data cube, where log(\nii/H$\alpha$)$<-0.2$ and log(\sii/H$\alpha$)$<-0.4$, which correspond to the innermost parts of the outflowing cone, indicated in orange in the right panels of Fig.~\ref{fig:cir_bpt}. After fitting and subtracting the continuum with \ppxf, we fitted the main emission lines with \mpfit, using one Gaussian component with the same velocity and velocity dispersion for each line. We applied the same procedure to obtain the stacked spectrum of the highest LILr in Circinus (log(\nii/H$\alpha$)$>0.1$ and log(\sii/H$\alpha$)$>0$), shown in the bottom panel of Fig.~\ref{fig:spectra}. In Table~\ref{table:1} the observed and dereddened fluxes of the fitted emission lines with respect to the H$\beta$ are reported for the two spectra. 
It can be seen that the Fe coronal lines (e.g. \fevi$\lambda5146$, \fexiv$\lambda5303$, \fevii$\lambda5721$, \fevii$\lambda6087$, \fex$\lambda6374$, highlighted in green) are brighter and are found almost exclusively  in the lowest LILr region. These forbidden transitions are peculiar because of their high-ionisation potential ($\geq 100$~eV), pointing to a scenario in which the innermost parts of the outflowing gas in this galaxy are in optically thin, highly ionised regions, possibly directly illuminated by the central ionising AGN.

Furthermore, the \nii$\lambda5755$ auroral line, which is  too faint to be detected in the single spaxel spectra,  is observed in the stacked spectrum in Fig.~\ref{fig:spectra}.
We can then determine the electron temperature ($T_e$) by exploiting the temperature-sensitive auroral-to-nebular line ratio of this particular ion. The atomic structure of \nii \, is such that auroral and nebular lines originate from excited states that are well spaced in energy, and thus their relative level populations depend heavily on electron temperature. From the \nii$\lambda\lambda$6548,84/\nii$\lambda$5755 line ratio, we obtain $T_e \sim 9.1\times 10^3 $~K. This value is lower than the few other measurements present in the literature, to the best of our knowledge, derived in outflows by \citet{villarmartin2014}, \citet{brusa2016}, \citet{perna2017} and \citet{nesvadba2018} (i.e. $T_e \approx 1.5 \times 10^4 $~K). This discrepancy is possibly due to the fact that these authors estimated the electron temperature on the basis of \oiii \, (i.e. \oiii$\lambda\lambda$4959,5007/\oiii$\lambda$4363), characterised by a higher ionisation potential with respect to \nii, which  mainly traces the innermost parts of the photoionised regions and can be associated with lower temperatures (\citealt{osterbrock2006,curti2017}; Perna et al. 2018 in preparation).

\begin{table*}
\caption{Circinus: observed and dereddened line fluxes ($F$ and $I$),   relative to H$\beta$ (H$\beta=1$),   and electron temperatures $T_e$ for the low LILr (log(\nii/H$\alpha$)~$<-0.2$ and log(\sii/H$\alpha$)~$<-0.4$) and high LILr (log(\nii/H$\alpha$)~$>0.1$ and log(\sii/H$\alpha$)~$>0$) regions, estimated by fitting the corresponding stacked spectra. Blank entries are undetected lines. H$\beta$ flux is in units of $10^{-16}$~erg s$^{-1}$ cm$^{-2}$, while $T_e$ is in Kelvin.}
\label{table:1}      
\centering          
\begin{tabular}{l c c c r }     
\hline\hline       
                                                                                                &   \multicolumn{2}{c}{Low LILr region} &   \multicolumn{2}{c}{High LILr region} \\ 
                                                                                                & F                                               & I                                     & F                                       & I   \\
\hline                    
   H$\beta$                                                                             & 1                                               & 1                                     &  1                                      & 1                                             \\  
   \oiii$\lambda$5007                                                                   &  11.14 $ \pm $0.04              &10.17$ \pm $0.03       & 4.69 $ \pm $0.02              & 4.12 $ \pm $0.02                       \\
   \fevi$\lambda$5146                                                           &  0.022 $ \pm $0.002             & 0.018 $ \pm $0.002    &  -                                    & -                                               \\
   \fexiv$\lambda$5303                                                          &  0.008$ \pm $0.002              & 0.006 $ \pm $0.002    & -                                     & -                                               \\
   \fevii$\lambda$5721                                                          &  0.210 $ \pm $0.002             & 0.128 $ \pm $0.001    & 0.098$ \pm $0.004             & 0.049 $ \pm $0.002               \\
   \nii$\lambda$5755                                                                    & 0.034 $ \pm $0.002              & 0.021 $ \pm $0.001    & 0.093$ \pm $0.004             & 0.045$ \pm $0.002                       \\
   \fevii$\lambda$6087                                                                  & 0.114$ \pm $0.002                       & 0.058 $ \pm $0.001    & 0.037$ \pm $0.004              & 0.014 $ \pm $0.001             \\
   \oi$\lambda$6300                                                                     &  0.359$ \pm $0.002              & 0.171 $ \pm $0.001    & 1.160$ \pm $0.006             & 0.402$ \pm $0.001                        \\   
   \siii$\lambda$6312                                                           & 0.069 $ \pm $0.002              & 0.033$ \pm $0.001             & -                                     & -                                               \\
   \fex$\lambda$6374                                                                    &  0.041 $ \pm $0.002             & 0.019 $ \pm $0.001    & -                                     & -                                               \\  
   \nii$\lambda$6584                                                                    & 4.23 $ \pm $0.01                        & 1.765$ \pm $0.002     & 10.98$ \pm $0.05               & 3.168 $ \pm $0.002             \\
   H$\alpha$                                                                            & 6.79 $ \pm $0.02                        & 2.86                          & 9.78 $ \pm $0.04                & 2.86                                  \\
   \sii$\lambda$6717                                                                    & 1.141 $ \pm $0.005              & 0.450 $ \pm $0.001    & 3.46$ \pm $0.02               & 0.921$ \pm $0.001                       \\
   \sii$\lambda$6731                                                                    &1.357 $ \pm $0.005                    & 0.532$ \pm $0.001     & 4.62 $ \pm $0.02               & 1.222$ \pm $0.001                      \\
   H$\beta$ flux                                                                                & 227.2  $ \pm $0.7                       & 5160$ \pm $40         & 160.9 $ \pm $0.7                & 13700 $ \pm $100                      \\
\hline                  
  $T_e$                                                                                         & \multicolumn{2}{c}{ $(9.1 \pm 0.2 ) \times10^3$}         & \multicolumn{2}{c}{$(9.8 \pm 0.2 ) \times10^3$}  \\
  \hline
  \hline
\end{tabular}
\end{table*}    

\subsection{Possible scenarios for the highest LILr}
\begin{figure*}
     \centering
           \begin{minipage}{2\columnwidth}
                \includegraphics[width=1.\columnwidth]{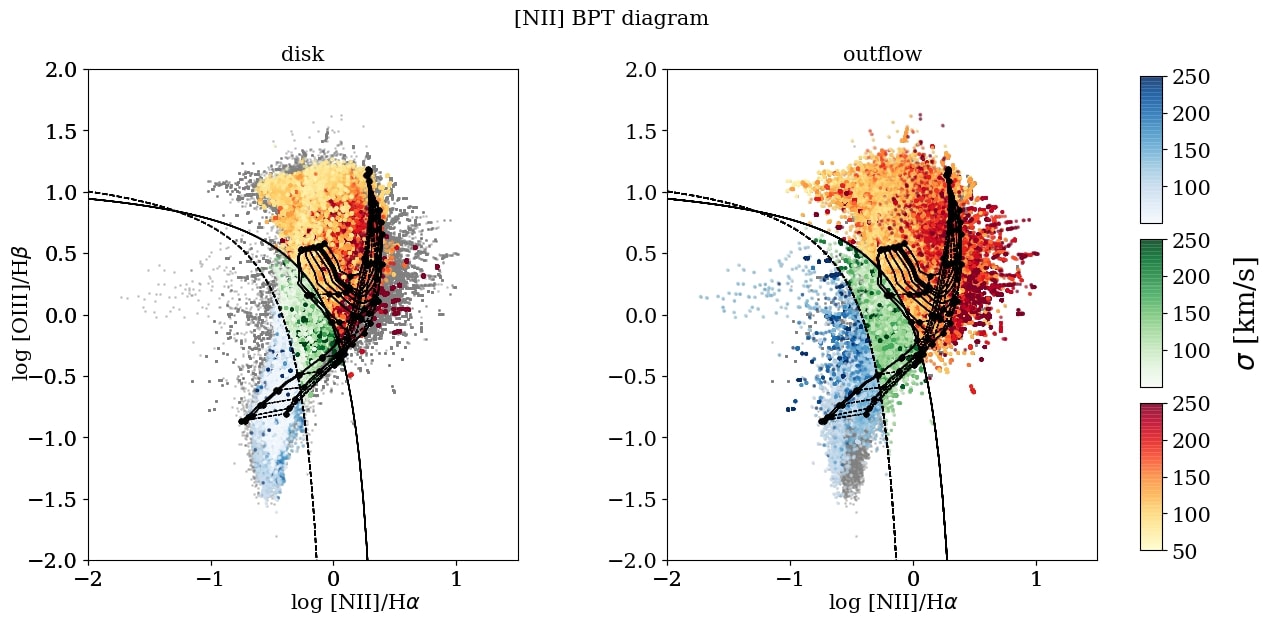}
        \end{minipage}
              \begin{minipage}{2\columnwidth}
                \includegraphics[width=1.\columnwidth]{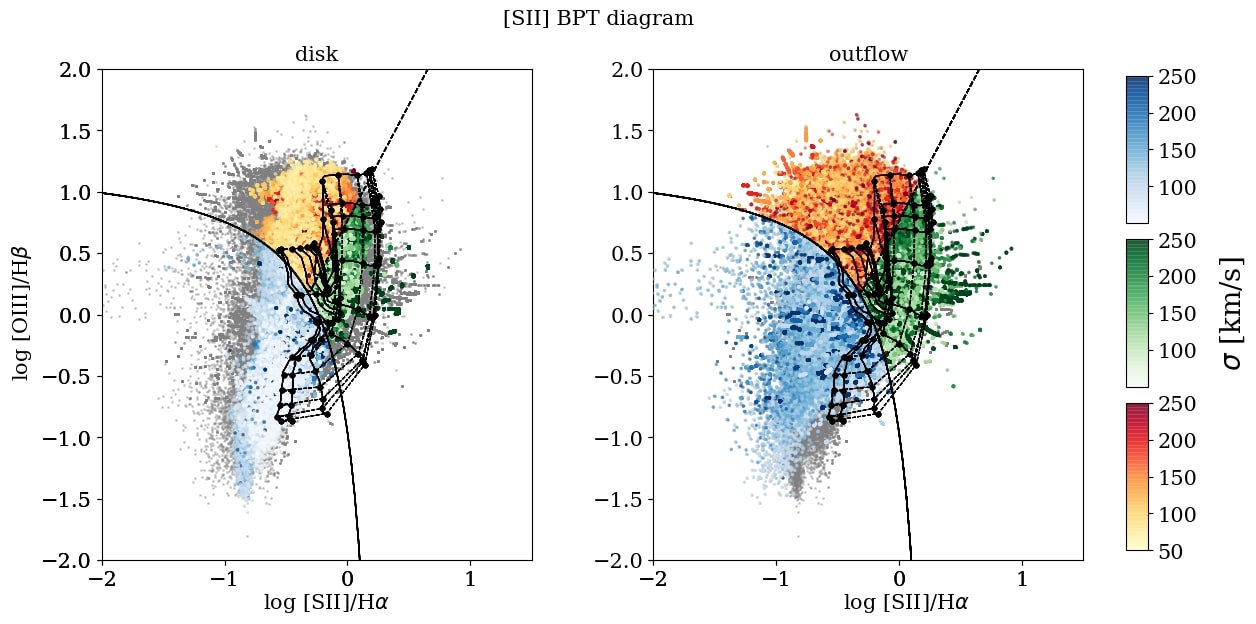}
        \end{minipage}
         \caption{\nii- and \sii-BPT diagrams for the disc and outflow components, as shown in Fig.~\ref{fig:bpt_total_mod1}, colour-coded as a function of the \oiii \, velocity dispersion (darker shades means higher $\sigma_{\rm \oiii}$). The grid of shock models taken from \citet{allen2008} comprises shock velocities in the range $v_s=100-1000$~km~s$^{-1}$ (horizontally increasing) and magnetic parameters $B/n^{1/2}=10^{-2}-10$~$\mu$G~cm$^{3/2}$ (vertically increasing). The black dotted line is the \citet{kewley2006} boundary between Seyferts and LI(N)ERs.}
                \label{fig:bpt_total_mod2}
   \end{figure*}

Figure~\ref{fig:bpt_total_mod2} shows the \nii- and \sii- BPT diagrams of all the MAGNUM galaxies (excluding NGC~1068 as before), coloured-coded as a function of the \oiii \, velocity dispersion $\sigma_{\rm \oiii}$, as reported in the corresponding colour bars (darker shades means higher $\sigma_{\rm \oiii}$). Only bins from the Voronoi-binned cubes in which the \oiii \, line flux has a S/N~$>3$ are shown. It can be clearly seen that the highest values of $\sigma_{\rm \oiii}$ correspond to the highest LILrs found in the outflowing gas. These  mainly trace the edges of the outflowing cones and/or the region perpendicular to it (see Sect.~\ref{sec:BPTs}). 
This correlation between velocity dispersion and emission-line ratio suggests that the kinematics and ionisation state are coupled, indicating that they may have the same physical origin. This trend has been already found in star-forming galaxies, U/LIRGs, and AGN  (e.g. \citealt{monrealibero2006, monrealibero2010, rich2011, ho2014, rich2015, mcelroy2015, perna2017}), and it was attributed to shocked gas components as photoionisation is not expected to cause such a trend (see \citealt{dopita1995} and the detailed discussion in \citealt{mcelroy2015}). 

Therefore, we compare the observed data with a grid of MAPPINGS III shock models\footnote{The values of the  magnetic field, $B$, and magnetic parameter, $B/{n_e}^{1/2}$,  were chosen by \citet{allen2008} so as to cover the extremes expected in the ISM. Moreover, the shock velocities taken into account are consistent with those measured from the gas kinematics (Venturi et al.  in preparation).}, taken from \citet{allen2008}, that comprises shock velocities in the range $v_s=100-1000$~km~s$^{-1}$ and magnetic parameters $B/{n_e}^{1/2}=10^{-2}-10$~$\mu$G~cm$^{3/2}$, taking into account solar abundances and a pre-shock density $n_e \sim100$~cm$^{-3}$. It can be clearly seen that shock models can reproduce the high LILrs (up to values of $\sim 0.3$) that the MB-IB dichotomy could not explain. 
However, the models fail to reproduce the highest LILrs, possibly suggesting even more extreme conditions in the outflowing gas (see also \citealt{perna2017}). We tested all the available \citet{allen2008} shock models, but found no improvement. Since this discrepancy is more significant in the \nii-BPT diagram, it may be partly related to a metallicity effect. Indeed, \nii/H$\alpha$ line-ratios should increase with higher metallicity due to the secondary nitrogen production (e.g. \citealt{alloin1979,considere2000,mallery2007}).

BPT diagrams colour-coded as a function of $\sigma_{\rm \oiii}$ for each galaxy can be seen in Fig.~\ref{afig:bptnii_sig_gal} and Fig.~\ref{afig:bptsii_sig_gal}, for \nii- and \sii-BPT diagrams, respectively. This correlation between \oiii \,velocity dispersion and high LILr is visible especially in NGC 4945, NGC~5643, and IC~5063. Also NGC~1068 shows this trend, even though it shows far higher values of \oiii \, velocity dispersion in all the outflowing components. The fact that this correlation is not found in Circinus does not necessarily exclude the shock scenario because, as stressed in \citet{mcelroy2015}, for idealised planar shock fronts we would only be observing the shock at a single velocity, implying that the observed shock velocity and velocity dispersion will not necessarily have similar values. 

From the \nii$\lambda\lambda$6548,84/\nii$\lambda$5755 line ratio, measured from the stacked spectrum shown in the bottom panel of Fig.~\ref{fig:spectra}, we obtain $T_e \sim 9.8 \times 10^3 $~K instead of the much higher temperature that we should observe if the bulk of emission were associated with shock excitation ($T_e 
> 10^5 $~K, \citealt{osterbrock2006}). A similar case has been found by \citet{perna2017}, who contend that a possible explanation could be that the shock accelerating the ISM gas is strongly cooled, so that the gas 
temperature rapidly returns to its pre-shock value (see also \citealt{king2014}).

Nevertheless, some works pointed out the inefficiency of shocks in producing line emission. For example, \citet{laor1998} showed that shocks can reprocess only $\sim10^{-6}$ of the rest mass to ionising radiation, with respect to a maximum conversion efficiency of $\sim 10^{-1}$ for the central continuum source, demonstrating that they can be a viable mechanism only in very low-luminosity sub-Eddington active galaxies, which are very inefficient in converting mass into radiation.
An alternative explanation to account for the regions producing the strongest LILrs can be given by a hard ionising radiation field resulting from radiation filtered by clumpy, ionised absorbers located tens of parsecs or less from the nucleus, possibly consistent with the torus dimension (e.g. \citealt{netzer2015}). This scenario is consistent with our finding of lower \siii/\sii \, line ratios (\siii/\sii~$<1$) in the regions characterised by the highest LILrs, with the exception of  NGC~1068.
Similar results have been obtained by \citet{kraemer2008}, who analysed HST images of NGC 4151 and found that the \oiii$\lambda\lambda$4959,5007/\oii$\lambda\lambda$3726,29 -- a proxy of the ionisation parameter analogous to \siii/\sii \, -- is lower near the edges of the ionisation bicone observed in the galaxy than along its axis, concluding that the structure of the NLR is due to filtering of the ionising radiation by ionised gas, and is consistent with disc-wind models. 
   
\section{Conclusions}\label{sec:conclusion}
In this paper, we explored the gas properties (e.g. density, ionisation parameter, reddening and source of ionisation) of the outflowing gas in the (E)NLR of the 9 nearby Seyfert galaxies that are part of the MAGNUM survey, all characterised by prominent conical or biconical outflows. Exploiting the very high spatial resolution of the optical integral field MUSE spectrograph at VLT, we were able to disentangle the outflow component from the disc component in order to analyse its peculiarities through spatially and kinematically resolved maps. To do this, we divided the main emission lines (H$\beta$, \oiii, \oi, H$\alpha$, \nii, \sii, \, and \siii) in velocity bins, associating the core of the lines (centred on the stellar velocity in each spaxel) with the disc, and the blueshifted and redshifted wings with the outflow. In the following we report our main results:

\begin{itemize}
\item In Sect.~\ref{sec:gas_properties}, we show that the outflow component is characterised by higher median densities and \siii/\sii \,ratio, which is a proxy for the ionisation parameter ($\langle A_V \rangle \sim0.9$, $\langle n_e \rangle \sim 250$~cm$^{-3}$, $\langle$log(\siii/\sii)$\rangle$~$\sim 0.16$) with respect to the gas in the disc, which instead is more affected by dust extinction ($ \langle A_V \rangle \sim1.75$, $ \langle n_e \rangle \sim 130$~cm$^{-3}$, $\langle$log(\siii/\sii)$\rangle$~$\sim -0.38$).  
Our median outflow density is lower with respect to what we found in the literature. However, calculating the median density weighting for the \sii \, line flux, we obtain higher values ($\langle n_e \rangle \sim 170$~cm$^{-3}$ and $\langle n_e \rangle \sim 815$~cm$^{-3}$, for disc and outflow, respectively). Therefore, many values of outflow density found in the literature could be biased towards higher $n_e$ because they are based only on the most luminous outflowing regions and are characterised by a higher S/N. 
\item In Sect.~\ref{sec:BPTs}, we analyse the ionisation state of the (E)NLR of MAGNUM galaxies, by making spatially and kinematically resolved BPT diagrams. We find that the AGN/LI(N)ER-dominated outflow is characterised by the lowest and highest values of low-ionisation line ratios (LILrs, log(\nii/H$\alpha$)~$ \sim -1$, log(\sii/H$\alpha$)~$ \sim -1$ and log(\nii/H$\alpha$)~$ \sim 0.5$, log(\sii/H$\alpha$)~$ \sim 0.5$, respectively), which are not observed in the disc. 
The lowest LILrs mainly come from the innermost regions of the outflowing cone, near the outflow axis, also characterised  by the highest \siii/\sii \, line ratios (\siii/\sii~$\sim 2$), meaning high excitation. On the other hand, the highest LILr appear to come from the edges of the outflowing cones and/or from the regions perpendicular to the axis of the outflow, typically characterised by high values of \oiii \, velocity dispersion ($\sigma_{\rm \oiii} > 200$~km/s).
\item In Sect.~\ref{sec:discussion}, we make a comparison of our spatially and kinematically resolved BPT diagrams with photoionisation and shock models. Specifically, we found that the matter and ionisation bounded (MB and IB) dichotomy first introduced by \citet{binette1996} could reproduce well most of the observed features. MB clouds, characterised by high-ionisation lines (e.g. high \oiii/H$\beta$) and high excitation (i.e. high \siii/\sii) could account for both the \nii/H$\alpha$, \sii/H$\alpha$ decrease and the \siii/\sii \, enhancement observed in the majority of MAGNUM galaxies. On the other hand, IB clouds, optically thick and characterised by a much lower excitation, could be responsible for higher \nii/H$\alpha$ and \sii/H$\alpha$ line ratios, found both in the outflowing and in the disc components. Shocks may explain the highest \nii/H$\alpha$ and \sii/H$\alpha$ line ratios and the \oiii \, velocity dispersion enhancement.
Consistently, we show the stacked spectrum of the region dominated by the lowest LILr of Circinus, characterised by the presence of coronal lines (e.g. \fevi$\lambda5146$, \fexiv$\lambda5303$, \fevii$\lambda5721$, \fevii$\lambda6087$, \fex$\lambda6374$), suggesting a very high excitation. 
\end{itemize}

We speculate that the gas in the outflowing cones of our galaxies is set up in clumpy clouds characterised by higher density and ionisation parameters with respect to the disc gas. The innermost regions of the cone are optically thin to the radiation, being characterised by high excitation, and are possibly directly heated by the central ionising AGN. The edges of the cones and the regions perpendicular to the outflow axis could instead be dominated by shock excitation probably because of the interaction between the outflowing gas and the ISM. Alternatively, these regions, generally characterised by low excitation (\siii/\sii~$<1$) could be impinged by an ionising radiation filtered by clumpy, ionised absorbers.
Although these conclusions apply to the MAGNUM sample as a whole, some details may differ in individual galaxies. As an example, NGC1068 is the only galaxy that shows an enhancement of the \siii/\sii \, line ratio in the regions characterised by the highest LILr.
The next step of our analysis will be a detailed modelling of each MAGNUM galaxy, also making  use of the photoionisation code \cloudy \, \citep{ferland2017} to better investigate the different scenarios that we propose to interpret our findings. 

\begin{acknowledgements}
These results are based on observations collected at the European Southern Observatory under ESO programme 094.B-0321(A). MM acknowledges the University of California Santa Cruz (UCSC) for its stimulating scientific environment and is grateful to Carlo Cannarozzo for inspiring conversations and advice. GC acknowledges the support by INAF/Frontiera through the `Progetti Premiali' funding scheme of the Italian Ministry of Education, University, and Research. 
GC, FM, AG, PT, and SZ have been supported by the INAF PRIN-SKA 2017 programme 1.05.01.88.04. CC, CF, and EN acknowledge funding from the European Union's Horizon 2020 research and innovation programme under the Marie Sk\l{odowska-Curie} grant agreement No 664931. RM acknowledges ERC Advanced Grant 695671 `QUENCH' and support by the Science and Technology Facilities Council (STFC). This research has made use of NASA Astrophysics Data System and of the NASA/IPAC Extragalactic Database (NED), which is operated by the Jet Propulsion Laboratory, California Institute of Technology, under contract with the National Aeronautics and Space Administration. 
\end{acknowledgements}

%
   \bibliographystyle{aa} 
   \bibliography{paper_magnum_mingozzi_2018} 

%
\begin{appendix} 
\section{Extinction maps}
\begin{figure*}
   \centering
\includegraphics[width=1.8\columnwidth]{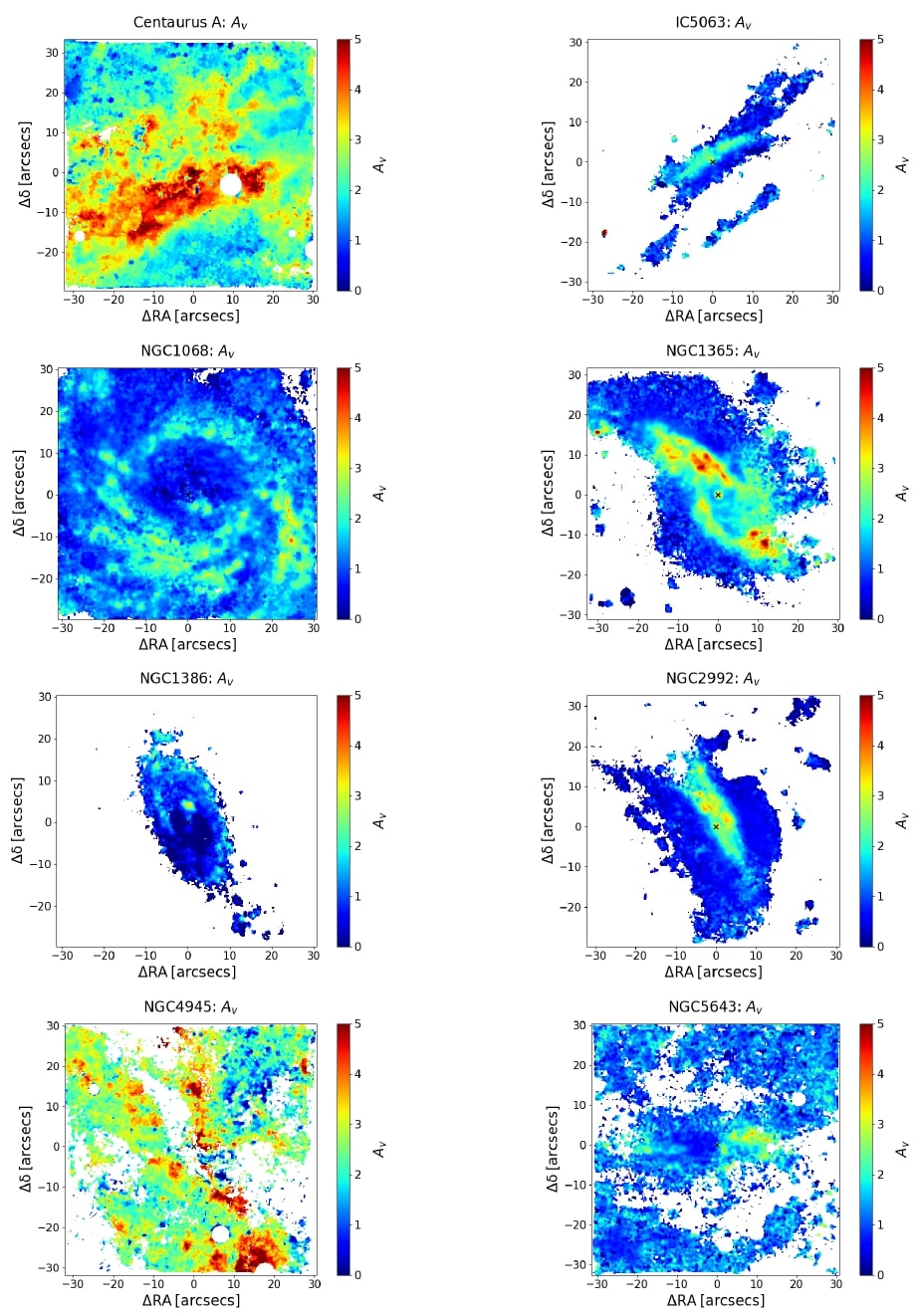}
	\caption{Maps of the total extinction in V band A$_{\rm V}$, obtained from the Balmer decrement H$\alpha$/H$\beta$, for Centaurus~A, IC~5063, NGC~1068, NGC~1365, NGC~1386, NGC~2992, NGC 4945 and NGC~5643, respectively. Only spaxels with H$\alpha$ and H$\beta$ SNR~$>5$ are shown.}
        	\label{afig:av_maps}
   \end{figure*}

We derive the dust extinction map for each galaxy, making use of the Balmer decrement H$\alpha$/H$\beta$, assuming a \citet{calzetti2000} attenuation law and a fixed temperature of $10^4$~K. The resulting extinction map for the MAGNUM galaxies Centaurus~A, IC~5063, NGC~1068, NGC~1365, NGC~1386, NGC~2992, NGC 4945, and NGC~5643 are shown in Fig.~\ref{afig:av_maps}. We show only the spaxels in which the H$\alpha$ and H$\beta$ lines are detected with S/N > 5. 

\section{Electron density maps}
\begin{figure*}
\centering
    \includegraphics[width=1.8\columnwidth]{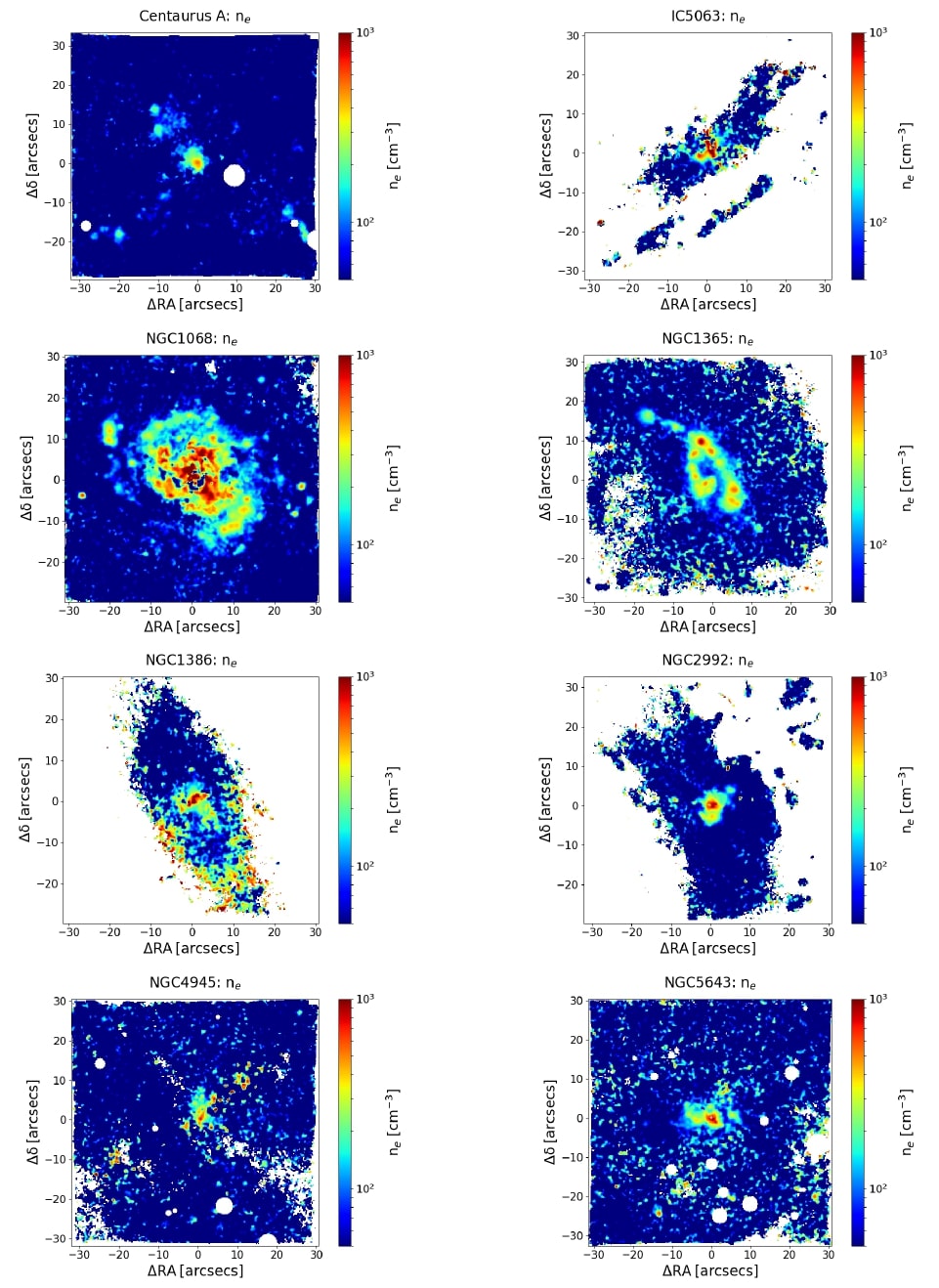}

	\caption{Maps of the total electron density $n_e$, measured from the \sii$\lambda$6717/\sii$\lambda$6731 ratio, for Centaurus~A, IC~5063, NGC~1068, NGC~1365, NGC~1386, NGC~2992, NGC 4945 and NGC~5643, respectively. Only spaxels with \sii$\lambda$6717 and \sii$\lambda$6731 SNR~$>5$ are shown.}
        	\label{afig:ne_maps}
   \end{figure*}

We have computed the \sii$\lambda\lambda$6717,31 line ratio in each spaxel where the \sii \, lines are detected with S/N > 5, and converted it to an electron density using the \citet{osterbrock2006} model, sensible to the density variations in the range $50$~cm$^{-3}<n_e<5000$~cm$^{-3}$, assuming a temperature of $10^4$~K. The resulting electron density maps of the MAGNUM galaxies Centaurus~A, IC~5063, NGC~1068, NGC~1365, NGC~1386, NGC~2992, NGC 4945, and NGC~5643 are shown in Fig.~\ref{afig:ne_maps}. 

\section{\siii/\sii \,maps}
\begin{figure*}
   \centering
\includegraphics[width=1.8\columnwidth]{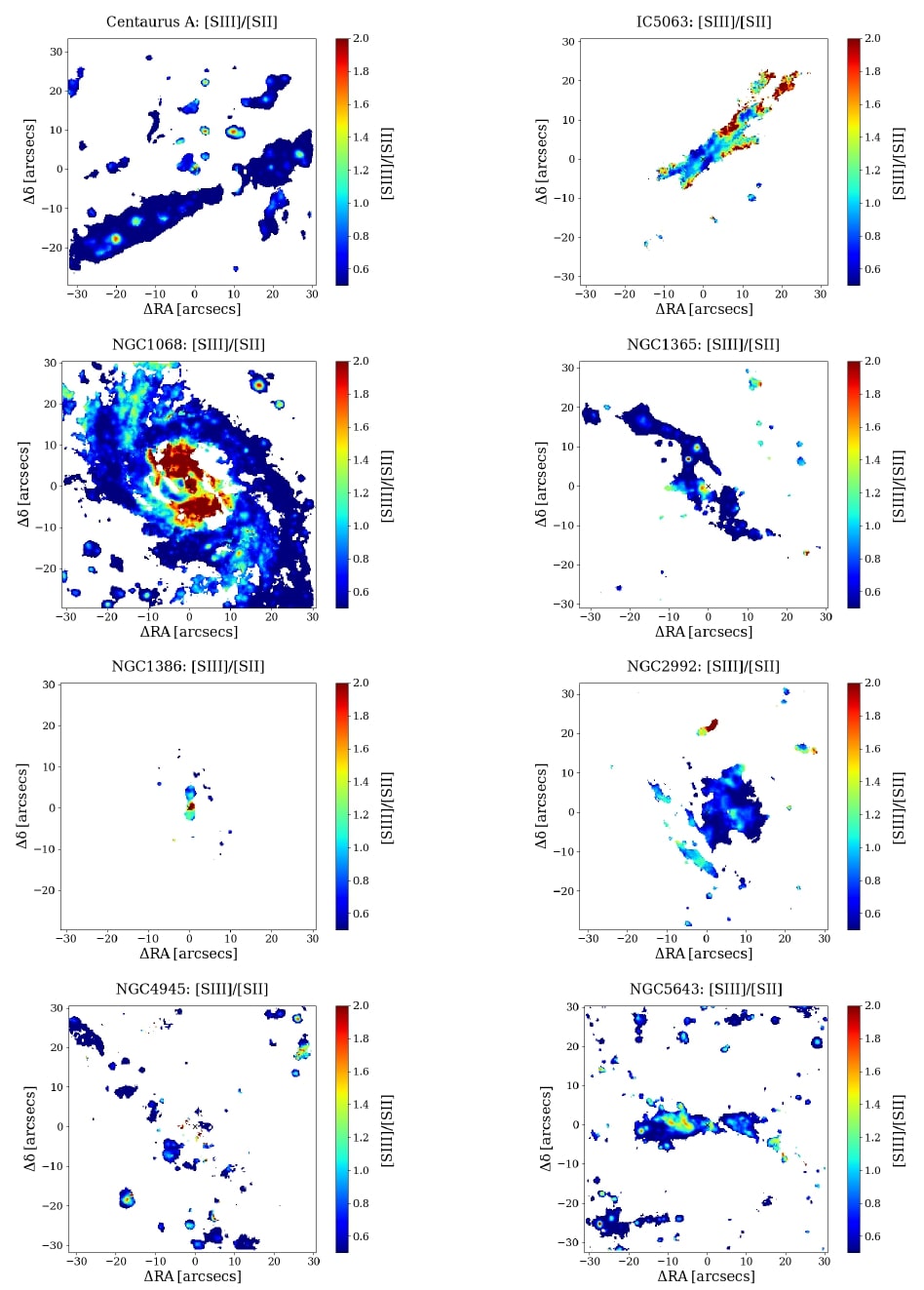}
	\caption{Maps of \siii$\lambda\lambda$9069,9532/\sii$\lambda\lambda$6717,31 ratio, proxy for the ionisation parameter, for Centaurus~A, IC~5063, NGC~1068, NGC~1365, NGC~1386, NGC~2992, NGC 4945 and NGC~5643, respectively. Only spaxels with \siii$\lambda$9069 and \sii$\lambda\lambda$6717,31 SNR~$>5$ are shown.}
        	\label{afig:ionu_maps}
   \end{figure*}

We computed the \siii$\lambda\lambda$9069,9532/\sii$\lambda\lambda$6717,31 line ratio in each MUSE spaxel where the line fluxes are detected with S/N~>~5 since it is a proxy for the ionisation parameter (e.g. \citealt{diaz2000}). The resulting maps for the MAGNUM galaxies Centaurus~A, IC~5063, NGC~1068, NGC~1365, NGC~1386, NGC~2992, NGC 4945, and NGC~5643 are shown in Fig.~\ref{afig:ionu_maps}. 

\section{Spatially and kinematically resolved BPT diagrams}
\begin{figure*}    
   \centering
    \includegraphics[width=1.85\columnwidth]{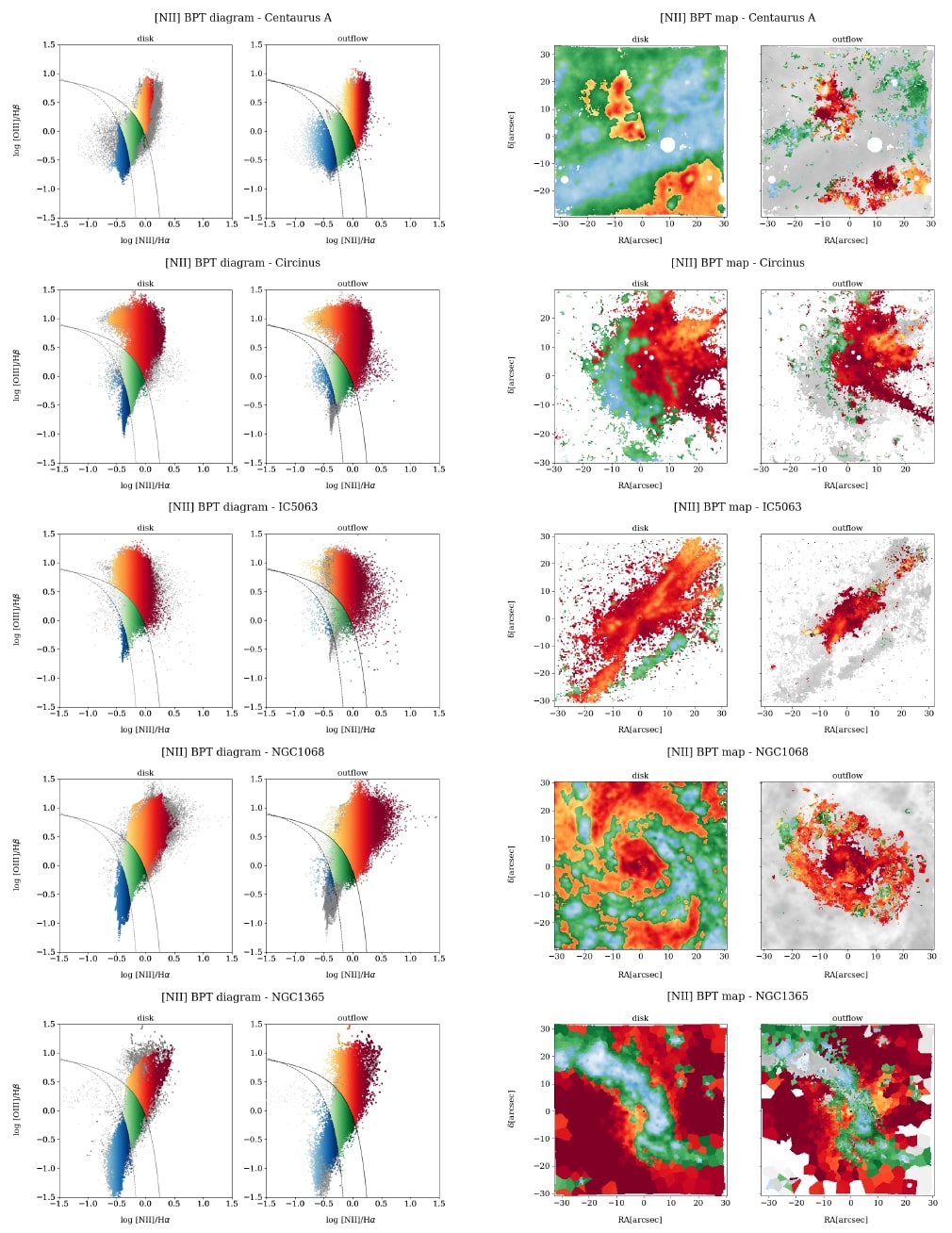}
      	\caption{Left panels: \nii-BPT diagrams for the disk and outflow components, on the left and the right respectively, of Centaurus~A, IC~5063, NGC~1068 and NGC~1365: shades of blue for SF, of green for composite regions, and of red for AGN-like ionising spectra, as a function of the \nii/H$\alpha$ line ratio (darker colour means higher \nii/H$\alpha$). The black dashed curve is the boundary between star-forming galaxies and AGN defined by \citet{kauffmann2003}, while the black solid one is the theoretical upper limit on SF line ratios found by \citet{kewley2001}. Right panels: \nii-BPT maps, coloured according to the BPT classification. In the background of all the pictures (black dots in the BPTs and shaded grey in the corresponding maps), are reported the disk and outflow component together. For each velocity bin, we select only the spaxels with a SNR~$>5$ for all the flux line ratios.}
        	\label{afig:bptnii_gal_1}
   \end{figure*}

\begin{figure*}
      \centering
    \includegraphics[width=1.85\columnwidth]{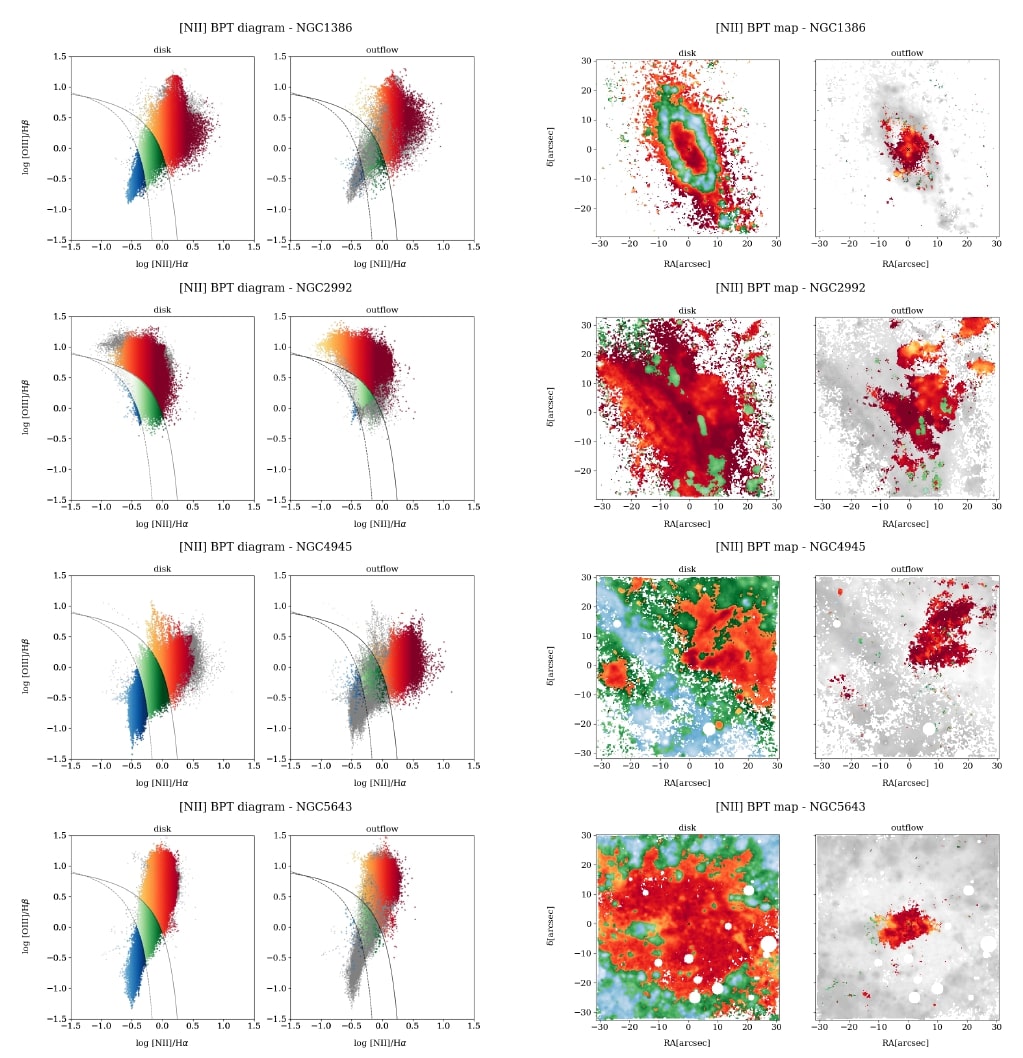}
      	\caption{Same as Fig.~\ref{afig:bptnii_gal_1} for NGC~1386, NGC~2992, NGC 4945 and NGC~5643.}
        	\label{afig:bptnii_gal_2}
   \end{figure*}

\begin{figure*}    
      \centering
    \includegraphics[width=1.85\columnwidth]{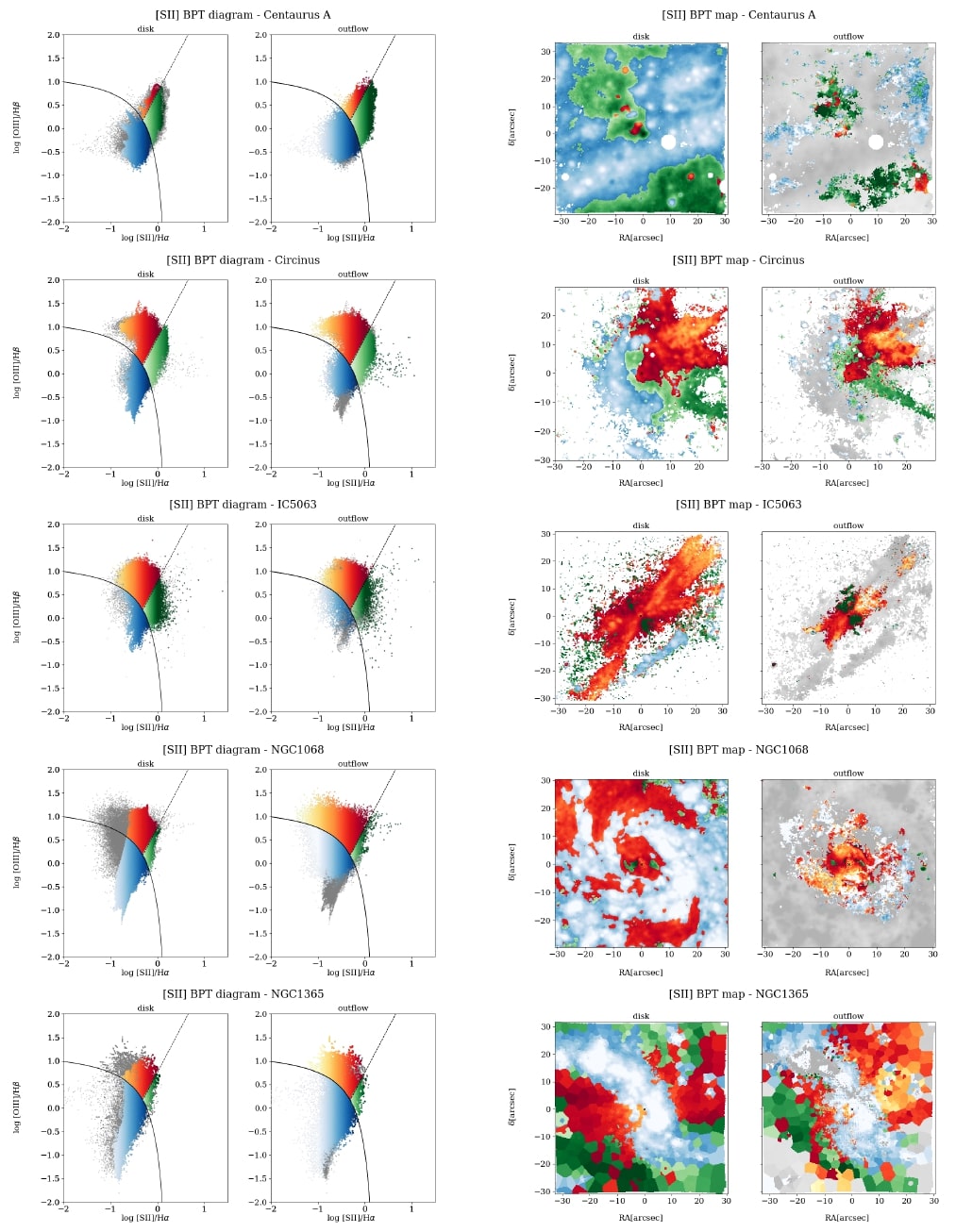}   
      	\caption{Left panels: \sii-BPT diagrams for the disk and outflow components, on the left and the right respectively, of Centaurus~A, IC~5063, NGC~1068 and NGC~1365: shades of blue for SF, of green for LI(N)ER, and of red for AGN-like ionising spectra, as a function of the \sii/H$\alpha$ line ratio (darker colour means higher \sii/H$\alpha$). The black solid curve is the theoretical upper limit on SF line ratios found by \citet{kewley2001}. The black dotted line is the \citet{kewley2006} boundary between Seyferts and LI(N)ERs. Right panels: \sii-BPT maps, coloured according to the BPT classification. In the background of all the pictures (black dots in the BPTs and shaded grey in the corresponding maps), are reported the disk and outflow component together. For each velocity bin, we select only the spaxels with a SNR~$>5$ for all the flux line ratios.}
        	\label{afig:bptsii_gal_1}
   \end{figure*}

\begin{figure*}  
   \centering
    \includegraphics[width=1.85\columnwidth]{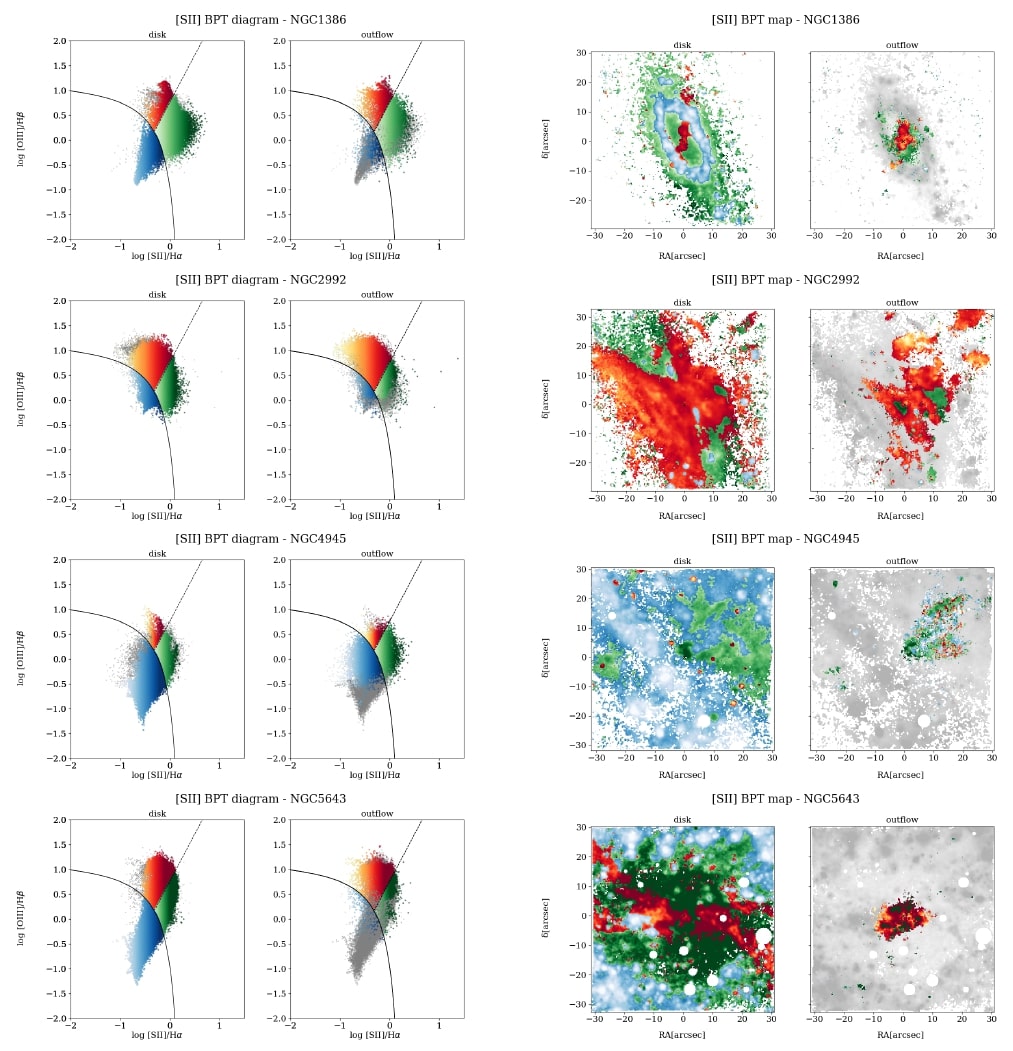}
      	\caption{Same as Fig.~\ref{afig:bptsii_gal_1} for NGC~1386, NGC~2992, NGC 4945 and NGC~5643.}
        	\label{afig:bptsii_gal_2}
   \end{figure*}

The left panels of Figs.~\ref{afig:bptnii_gal_1}, \ref{afig:bptnii_gal_2}, and Figs.~\ref{afig:bptsii_gal_1}, \ref{afig:bptsii_gal_2} show the \nii- and \sii-BPT diagrams for the disc and outflow components of Centaurus~A, Circinus, IC~5063, NGC~1068, NGC~1365, NGC~1386, NGC~2992, NGC 4945, and NGC~5643, respectively. The dashed curve is the boundary between star-forming galaxies and AGN defined by \citet{kauffmann2003}, while the solid curve is the theoretical upper limit on SF line ratios found by \citet{kewley2001}. The dotted line, instead, is the boundary between Seyferts and LI(N)ERs introduced by \citet{kewley2006}. The dominant source of ionisation is   colour-coded: blue for SF, green for intermediate regions in the \nii-BPT and LI(N)ER in the \sii-BPT, and red for AGN-like ionised spectra. The colours are shaded as a function of the x-axis line ratios (darker shades means higher x-axis line ratios). The corresponding position on the map of the galaxy, colour-coded according to  the different source of ionisation, is shown in the right panels. In the background of all the pictures (black dots in the BPTs and shaded grey in the corresponding maps) are reported the disc and outflow components together to allow a better visual comparison. 

\begin{figure*}
   \centering
    \includegraphics[width=1.9\columnwidth]{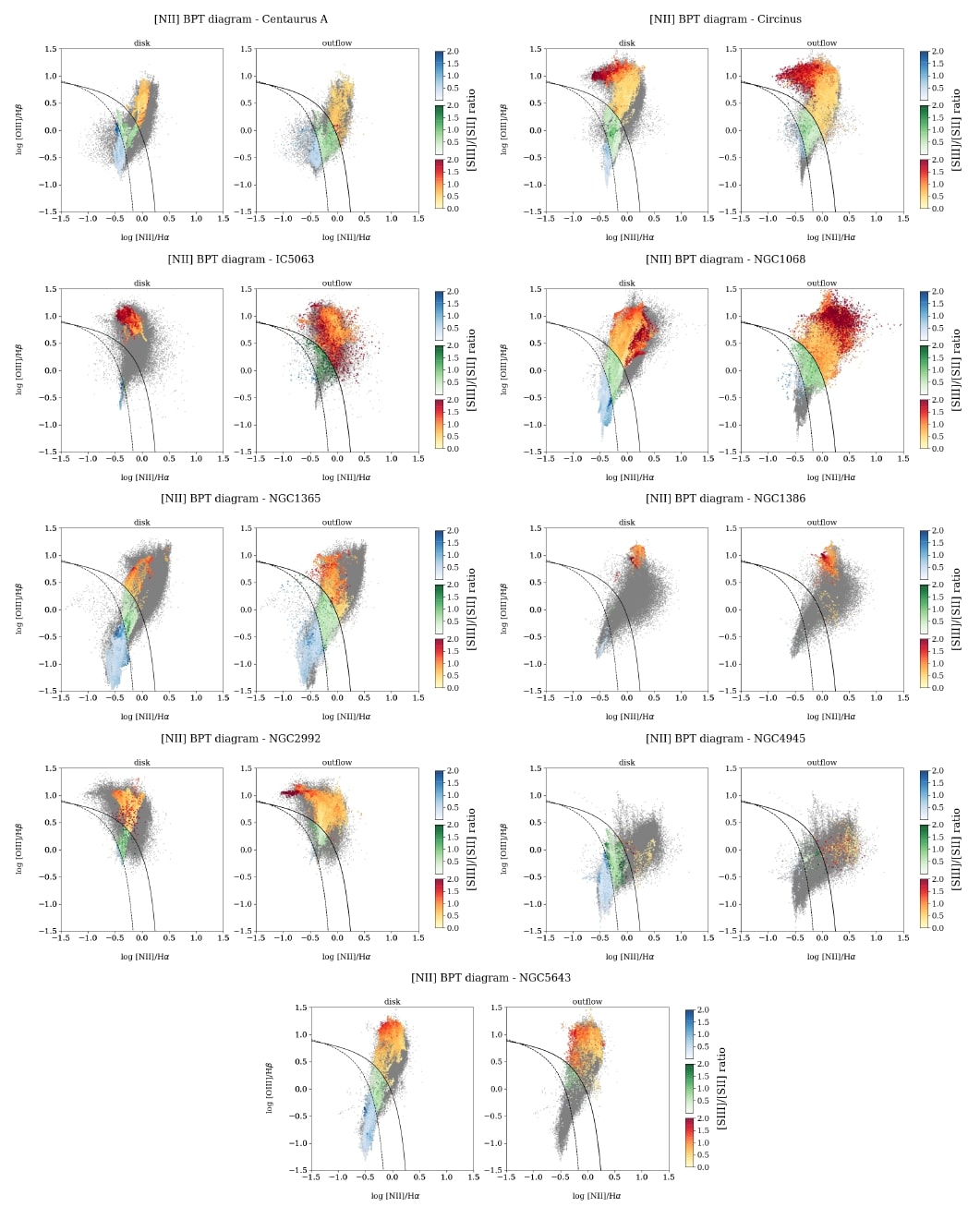}
      	\caption{\nii-BPT diagrams for the disk and outflow components for all the MAGNUM galaxies, colour coded as a function of the \siii/\sii \, line ratio (darker colour means higher \siii/\sii).}
        	\label{afig:bptnii_ionu_gal}	
   \end{figure*}

\begin{figure*}
      \centering
    \includegraphics[width=1.9\columnwidth]{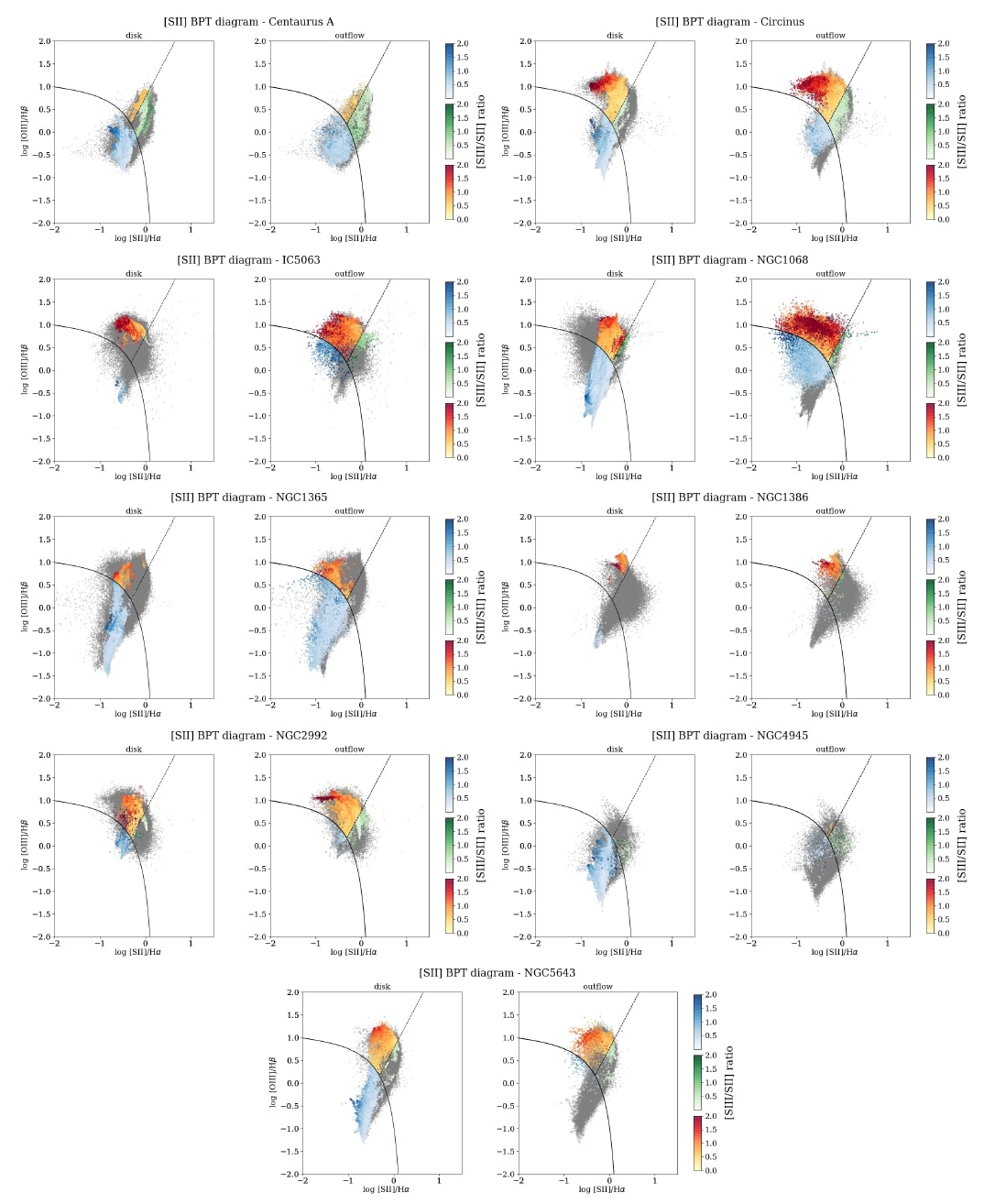}
      	\caption{\sii-BPT diagram for the disk and outflow components for all the MAGNUM galaxies, colour coded as a function of the \siii/\sii \, line ratio (darker colour means higher \siii/\sii).}
        	\label{afig:bptsii_ionu_gal}
   \end{figure*}

\begin{figure*}
   \centering
    \includegraphics[width=1.9\columnwidth]{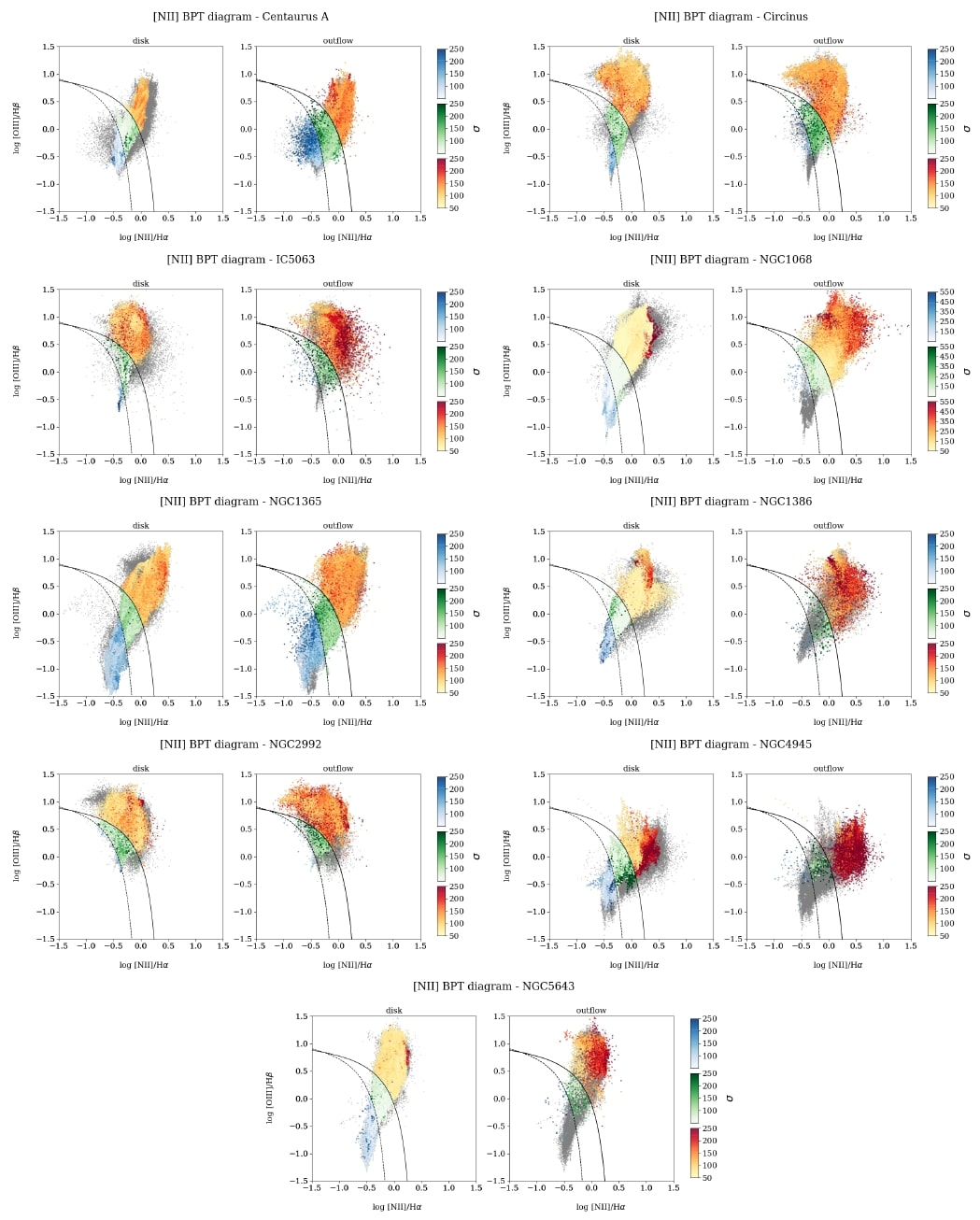}   
      	\caption{\nii-BPT diagram for the disk and outflow components for all the MAGNUM galaxies, colour coded as a function of the \oiii \, velocity dispersion $\sigma_{\rm \oiii}$ (darker colour means higher $\sigma_{\rm \oiii}$).}
        	\label{afig:bptnii_sig_gal}
   \end{figure*}

\begin{figure*}
   \centering
    \includegraphics[width=1.9\columnwidth]{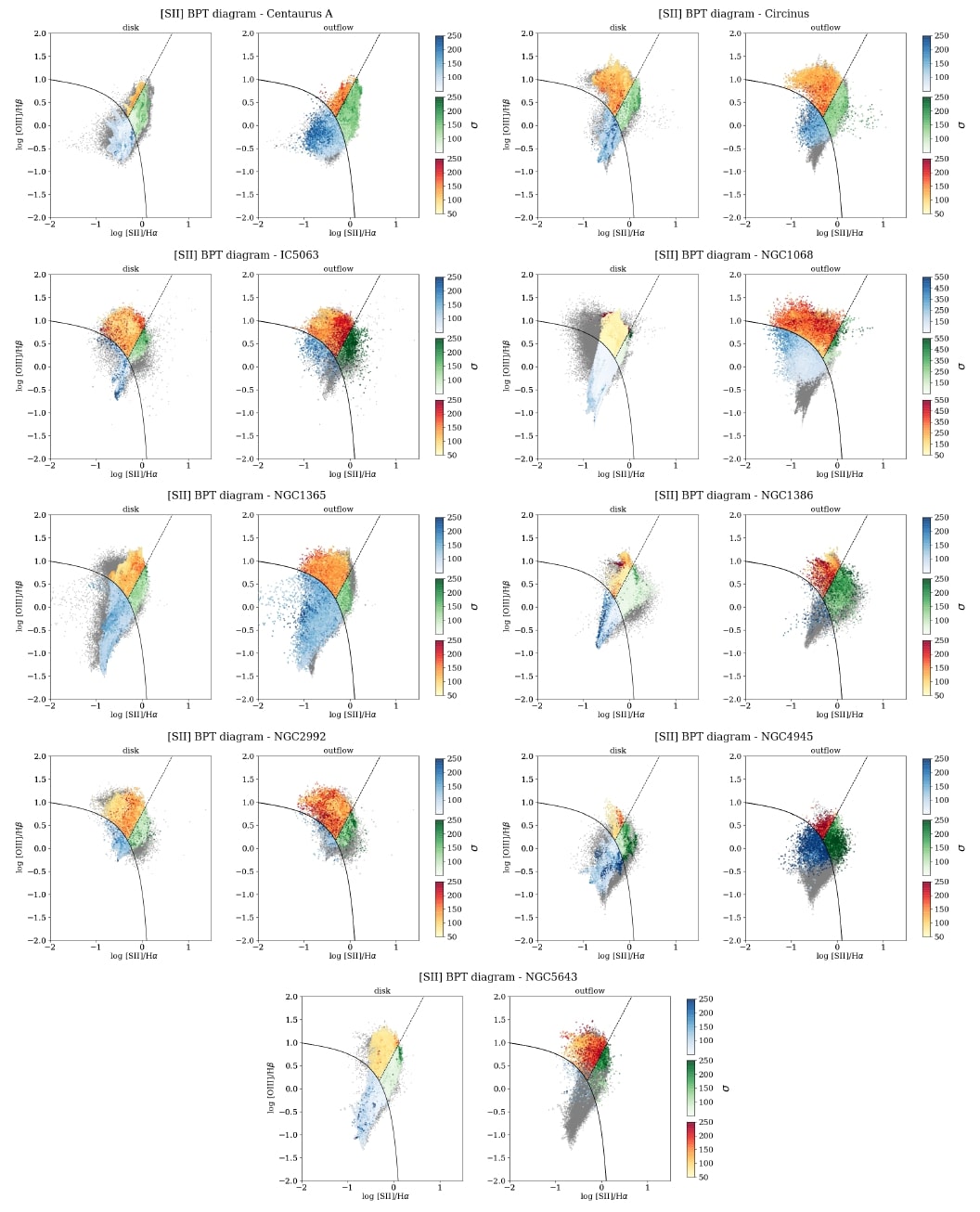}
      	\caption{\sii-BPT diagram for the disk and outflow components for all the MAGNUM galaxies, colour coded as a function of the \oiii \, velocity dispersion $\sigma_{\rm \oiii}$ (darker colour means higher $\sigma_{\rm \oiii}$).}
        	\label{afig:bptsii_sig_gal}
   \end{figure*}

Moreover, Figs.~\ref{afig:bptnii_ionu_gal} and \ref{afig:bptsii_ionu_gal} show the \nii- and \sii-BPT diagrams colour-coded as a function of the \siii/\sii \, line ratio, which is a proxy of the ionisation parameter (darker shades means higher \siii/\sii \, line ratios).
Finally, Figs.~\ref{afig:bptnii_sig_gal} and \ref{afig:bptsii_sig_gal} show the \nii- and \sii-BPT diagrams colour-coded as a function of the \oiii \, velocity dispersion $\sigma_{\rm \oiii}$ (darker shades means higher $\sigma_{\rm \oiii}$).

For each velocity bin, we selected only the spaxels with a S/N~$>5$ on each emission line involved in the given diagram, computed by dividing the flux integrated in the velocity channels considered by the corresponding noise.

For NGC~1365, we show the BPT diagrams obtained from the Voronoi-binned cube in order to detect the biconical outflow, which would be too weak to be detected in H$\beta$ otherwise.

\end{appendix}


\end{document}